\documentclass[journal]{IEEEtran}

\newcounter{mytempeqncnt}

\ifCLASSINFOpdf
  \usepackage[pdftex]{graphicx}
  \graphicspath{{./}{Figs/}}
  \DeclareGraphicsExtensions{.pdf,.jpeg,.png}
\else
\fi
\usepackage{amssymb}
\usepackage{amsmath}
\usepackage{mathtools}
\usepackage{bigdelim}
\newcommand{\MeijerG}[7]{G^{#1,#2}_{#3,#4} \left( \begin{smallmatrix} #5 \\ #6 \end{smallmatrix} \middle\vert #7 \right) }

\begin{document}
\title{Spectrum Sensing Via Reconfigurable Antennas: Fundamental Limits and Potential Gains}
\author{Ahmed~M.~Alaa,~\IEEEmembership{Student Member,~IEEE}, Mahmoud~H.~Ismail,~\IEEEmembership{Member,~IEEE} and~Hazim~Tawfik
\thanks{The authors are with the Department
of Electronics and Electrical Communications Engineering, Cairo University, Gizah,
12613, Egypt (e-mail: \{aalaa, mismail, htawfik\}@eece.cu.edu.eg).}
\thanks{Manuscript received XXXX XX, 201X; revised XXXX XX, 201X.}}

\markboth{XXXX,~Vol.~XX, No.~X, XXXX~201X}
{Alaa \MakeLowercase{\textit{et al.}}: Spectrum Sensing Via Reconfigurable Antennas: Fundamental Limits and Potential Gains}

\maketitle
\begin{abstract}
We propose a novel paradigm for spectrum sensing in cognitive radio networks that provides diversity and capacity benefits using a single antenna at the Secondary User (SU) receiver. The proposed scheme is based on a \textit{reconfigurable antenna}: an antenna that is capable of altering its radiation characteristics by changing its geometric configuration. Each configuration is designated as an antenna \textit{mode} or \textit{state} and corresponds to a distinct channel realization. Based on an abstract model for the reconfigurable antenna, we tackle two different settings for the cognitive radio problem and present fundamental limits on the achievable diversity and throughput gains. First, we explore the ``\textit{to cooperate or not to cooperate}" tradeoff between the diversity and coding gains in conventional cooperative and non-cooperative spectrum sensing schemes, showing that cooperation is not always beneficial. Based on this analysis, we propose two sensing schemes based on reconfigurable antennas that we term as \textit{state switching} and \textit{state selection}. It is shown that each of these schemes outperform both cooperative and non-cooperative spectrum sensing under a global energy constraint. Next, we study the ``\textit{sensing-throughput}" trade-off, and demonstrate that using reconfigurable antennas, the optimal sensing time is reduced allowing for a longer transmission time, and thus better throughput. Moreover, state selection can be applied to boost the capacity of SU transmission.

\end{abstract}

\begin{keywords}
cognitive radio; cooperative spectrum sensing; diversity; ergodic capacity; reconfigurable antennas; spectrum sensing
\end{keywords}

\IEEEpeerreviewmaketitle
\section{Introduction}
\IEEEPARstart{C}{ognitive} Radio (CR) is a promising technology offering a significant enhancement in wireless systems spectrum efficiency via dynamic spectrum access \cite{1}. In a CR network, unlicensed secondary users (SUs) can opportunistically occupy the unused spectrum allocated to a licensed primary user (PU). This is achieved by means of PU signal detection. Detection of PU signal entails sensing the spectrum occupied by the licensed user in a continuous manner. Thus, the process of \textit{spectrum sensing} is mandatory for a CR system as it helps preserving the Quality-of-Service (QoS) experienced by the licensed PU. Energy detection (ED) is one of the simplest spectrum sensing techniques as it can be implemented using simple hardware and does not require Channel State Information (CSI) at the SU receiver \cite{2}--\cite{3}. Generally, the performance of a spectrum sensing technique severely degrades in slow fading channels. To combat this effect, Cooperative Spectrum Sensing (CSS) schemes have been proposed to take advantage of the spatial diversity in wireless channels \cite{4}--\cite{6}. In CSS, hard or soft decisions from different CR users are combined to make a global decision at a central unit known as the \textit{Fusion Center} (FC). CSS has been widely accepted in the literature as a realizable technique for extracting spatial diversity. The other alternative would be using multiple antennas, which is constrained by the space limitation in SU recievers \cite{6}--\cite{14}.

\subsection{Background and Motivation}
Although CSS achieves a diversity gain that is equal to the number of cooperating users, it encounters a significant cooperation overhead: several decisions taken at SU terminals have to be fed back to the FC via a dedicated reporting channel \cite{5}; global information (including the number of cooperating SU terminals) must be provided to each SU in order to calculate the optimal detection threshold \cite{6}; hard decisions taken locally at each SU cause loss of information, which degrades the performance at low signal-to-noise ratio (SNR) \cite{7}; and finally, the existence of multiple SUs is not always guaranteed. In addition, in this work, we show that there exists a trade-off between the coding gain and the diversity order achieved in both cooperative and non-cooperative schemes, and demonstrate that cooperation is actually not beneficial in the low SNR regime. Motivated by these disadvantages, we tackle the following question: \textit{can we dispense with secondary users cooperation and still achieve an arbitrary diversity gain?} To answer this question, we propose a novel spectrum sensing scheme that can indeed achieve an arbitrary diversity order for a single SU and still uses a single antenna. The scheme is based on the usage of \textit{reconfigurable antennas}; a class of antennas capable of changing its geometry, hence changing the current distribution over the volume of the antenna and thus altering one of its propagation characteristics: operating frequency, polarization or radiation pattern. Each geometrical configuration thus leads to a different mode of operation leading to different realizations of the perceived wireless channel. Switching between various antenna modes could be done using microelectromechanical (MEMS) switches \cite{15}, nanoelectromechanical switches (NEMS), or solid state switches \cite{22}.

In \cite{15}, the concept of an electrically reconfigurable antenna was first introduced based on RF MEMS switches. Many research efforts followed this concept and proposed actual designs for antennas that can alter their geometric configuration \cite{16}--\cite{20}. The usage of reconfigurable antennas in wireless communications was studied in various contexts. For example, based on an abstract conceptual model, diversity benefits of reconfigurable antennas in MIMO systems were discussed in \cite{19}. Also, in \cite{21}, a new class of space-time codes, termed as \textit{state-space-time} codes was introduced, where it was shown that reconfigurable antennas can offer diversity benefits but has no impact on the achieved degrees of freedom. Moreover, reconfigurable antennas were employed in the context of interference alignment in \cite{22}, where desirable channel fluctuations were created by switching the antenna modes over time.

\subsection{Summary of Contributions}
In this paper, we propose a single user CR system that employs a reconfigurable antenna at the SU transceivers. By switching the antenna \textit{radiation states} over time, we can manipulate the wireless channel thus creating artificial channel fluctuations that turn a slow fading channel into a fast fading one. Capitalizing on this property, we show that we can dispense with the spatial diversity achieved through cooperation without encountering any degradation in the sensing performance. Besides, the proposed scheme has the following advantages: 1) the full coding and diversity gains are captured at any SNR, 2) the space limitation problem that inhibits the usage of multiple antennas is solved by using a single compact antenna, 3) unlike multiple antenna systems, only one RF chain is needed, 4) the availability of CSI at the SU can be used to even boost the achieved coding gain, and 5) diversity is achieved with no cooperation overhead, which usually involves setting up a dedicated reporting channel; feeding back information from the FC to the SU terminals; and maintaining synchronization between the SU devices.

Another approach for sensing using reconfigurable antennas is to select the ``best" state instead of randomly switching among various states. When the CSI is available at the SU, the receiver can select the state that offers the strongest channel gain. Therefore, in addition to the previously stated advantages, state selection offers an additional SNR gain, that we term as the \textit{selection gain}. Based on a comprehensive diversity analysis, we obtain the achievable diversity orders in the conventional and proposed schemes as a function of the detection threshold based on Neyman-Pearson (NP) and Bayes tests.

While there exists many antenna switching techniques with different ranges of switching delays \cite{17}, some classes of switching devices, such as those based on mechanical switches, may exhibit significant switching delays that may affect the performance of the proposed schemes. Thus, we quantify the impact of an arbitrary switching delay on the performance of the proposed schemes in both the NP and Bayesian tests. 

Moreover, we revisit a well known trade-off in CR systems, which is the ``Sensing-throughput trade-off". In a frame-structured CR system, each frame duration is divided into sensing and transmission periods. An optimal sensing time that compromises between the detection performance and the achieved throughput was calculated in \cite{23}--\cite{25}. We show that using reconfigurable antennas, and given a constraint on the PU detection probability, the SU throughput is improved as a longer period of the frame can be dedicated to transmission rather than sensing, in addition to the reduction of the false alarm probability, which means better channel utilization.

Finally, we show that reconfigurable antennas are not only beneficial in the sensing phase, but can also offer significant capacity gains in the transmission phase (when the SU accesses the channel). To that end, we obtain closed-form expressions for the average transmission capacity using state selection, and taking into consideration the impact of switching delay.

The rest of the paper is organized as follows: Section II presents the signal model adopted in the spectrum sensing problem and relevant derivations for the false alarm and missed detection probabilities. In Section III, we discuss the ``to cooperate or not to cooperate" tradeoff, identifying the drawbacks of the cooperative scheme. Spectrum sensing via reconfigurable antennas is introduced in Section IV, and the diversity orders obtained in sensing based on NP and Bayes criterion are derived. In section V, the impact of reconfigurable antennas on the sensing-throughput tradeoff is studied, showing the achievable throughput gains. In addition, the gains achieved in SU transmission and the optimal switching strategy are analyzed. Finally, we draw our conclusions in Section VI.

\section{Spectrum Sensing Signal Model}
\subsection{System Model and Notations}
In this section, we formulate the spectrum sensing problem for the conventional and proposed schemes, and clarify the notations of \textit{diversity order} and \textit{coding gain}.

The diversity order $d_{*}$ for a performance metric $P_{*}$ with an average SNR of $\overline{\gamma}$ is defined as \cite{6}
\[d_{*} = -\lim_{\overline{\gamma}\to \infty} \frac{\log P_{*}}{\log \overline{\gamma}}.\] The performance metric $P_{*}$ usually represents either the probability of error, the false alarm probability or the missed detection probability. The metric $P_{*}$ corresponds to the missed detection probability $P_{md}$ in the NP optimization problem, and corresponds to the error probability $P_{e}$ if the optimization problem adopts the Bayesian criterion. As for the coding gain, it is defined as the multiplication factor of the average SNR in $P_{*}$ as $\overline{\gamma}$ tends to infinity. Thus, if $P_{*}$ $\asymp$ $\frac{1}{(A \overline{\gamma})^d}$ as $\overline{\gamma}$ $\mapsto$ $\infty$, the coding gain is given by $A$ and the diversity order is $d$, where $\asymp$ denotes asymptotic equality. The diversity order affects the slope of the $P_{*}$ curve when plotted versus the average SNR (in dB), while the coding gain shifts the $P_{*}$ curve along the SNR curve. In spectrum sensing using energy detection, the coding gain is indeed sensitive to the average energy involved in detection. Hence, the average energy can be used to quantify the shift of the $P_{*}$ curve. Without loss of generality, we are interested in evaluating the asymptotic missed detection and error probabilities at high SNR only in order to obtain the diversity order and coding gain using the previous definitions. It is important to note, however, that both gains characterize the performance for all ranges of SNR.

Now, it is required to compare the detection performance of non-cooperative sensing, cooperative sensing, and reconfigurable antenna based schemes. Hereunder, we present the system model for the three schemes under study.

\subsubsection{Non-cooperative scheme}
A conventional non-cooperative spectrum sensing scheme involves one SU that observes $M$ samples for spectrum sensing. According to the sampling theorem, for a sensing period of $T$ and a signal with bandwidth $W$, the number of samples is $M$ = 2 $TW$ \cite{26}. It is assumed that the instantaneous SNR is $\gamma$ and the primary signal $i^{t h}$ sample is $S_{i}$ $\sim$ $\mathcal{CN}(0, 1)$ \cite{7}, where $\mathcal{CN}(\mu, \sigma^2)$ denotes the complex Gaussian distribution with mean $\mu$ and variance $\sigma^2$. The additive white noise is $n_{i}$ $\sim$ $\mathcal{CN}(0, 1)$. Thus, the $i^{t h}$ sample received at the SU receiver is a binary hypothesis given by \cite{7}
\begin{equation}
\label{eqn_example}
   r_{i} = \left\{
     \begin{array}{lr}
       n_{i} \sim \mathcal{CN}(0, 1), & \ \mathcal{H}_{o} \\
       \sqrt{\gamma} \hspace{1 mm} S_{i} + n_{i}  \sim \mathcal{CN}(0, 1 + \gamma), & \ \mathcal{H}_{1}
     \end{array}
   \right.
\end{equation}
where $\mathcal{H}_{o}$ denotes the absence of the PU, while $\mathcal{H}_{1}$ denotes the presence of the PU. After applying such signal to an energy detector, the resulting test statistic is $Y = \sum_{i=1}^{M} |r_{i}|^{2}$, which follows a central chi-squared distribution for both $\mathcal{H}_{o}$ and $\mathcal{H}_{1}$. The false alarm and detection probabilities are given by \cite{7}
\[P_{F}(M,\lambda) = P(Y > \lambda | \mathcal{H}_{o}) = \frac{\Gamma (M,\frac{\lambda}{2})}{\Gamma(M)},\]
and
\begin{equation}
\label{eqn_example}
P_{D}(M,\lambda,\gamma) = P(Y \leq \lambda | \mathcal{H}_{1}, \gamma) = \frac{\Gamma \left(M,\frac{\lambda}{2(1 + \gamma)}\right)}{\Gamma(M)},
\end{equation}
where $\lambda$ is the detection threshold, $\Gamma(.,.)$ is the upper incomplete gamma function, and $\Gamma(.)$ is the gamma function. We assume Rayleigh fading with an average SNR of $\overline{\gamma}$ and that the instantaneous SNR is constant over the $M$ observed samples (slow fading). Different observations perceive different SNR values. The SNR varies according to the exponential probability density function (pdf)
\begin{equation}
\label{eqn_example}
f_{\gamma}(\gamma) = \frac{1}{\overline{\gamma}} e^{-\frac{\gamma}{\overline{\gamma}}}, \gamma \geq 0.
\end{equation}

Because the detection probability is a function of the slow fading channel gain, we obtain the average detection probability as
\begin{equation}
\label{integ_mej}
\overline{P}_{D} = \int_0^{\infty} \frac{\Gamma \left(M,\frac{\lambda}{2(1 + \gamma)}\right)}{\Gamma\left(M\right)} \frac{1}{\overline{\gamma}} e^{-\frac{\gamma}{\overline{\gamma}}}  \,d\gamma.
\end{equation}
In order to evaluate the average detection probability, we can rewrite the integrands in (\ref{integ_mej}) in terms of the Meijer-G function $\MeijerG{m}{n}{p}{q}{a_1,\ldots,a_p}{b_1,\ldots,b_q}{z}$ [27, Sec. 7.8] as
\[\Gamma \left(M,\frac{\lambda}{2(1+\gamma)}\right) = \MeijerG{2}{0}{1}{2}{1}{M, \hspace{0.5 mm} 0}{\frac{\lambda}{2(1+\gamma)}},\]
and
\[e^{-\frac{\gamma}{\overline{\gamma}}} = \MeijerG{1}{0}{0}{1}{-}{0}{ \frac{\gamma}{\overline{\gamma}}}.\]
The Meijer-G representation allows us to write the integral in (\ref{integ_mej}) as
\begin{equation}
\overline{P}_{D} = \int_0^{\infty} \MeijerG{1}{0}{0}{1}{-}{0}{ \frac{\gamma}{\overline{\gamma}}} \MeijerG{2}{0}{1}{2}{1}{M, \hspace{0.5 mm} 0}{\frac{\lambda}{2(1+\gamma)}} d\gamma.
\end{equation}
 With the aid of [27, Eq. 7.811.1], the integral is approximated at high SNR as
\begin{equation}
\overline{P}_{D} \approx \frac{\lambda e^{\frac{1}{\overline{\gamma}}}}{2 \hspace{0.5 mm} \overline{\gamma} \hspace{0.5 mm} \Gamma(M)}  \hspace{0.5 mm}  \MeijerG{3}{0}{1}{3}{0}{M-1, \hspace{0.5 mm} -1, \hspace{0.5 mm} 0}{\frac{\lambda}{2 \hspace{0.5 mm} \overline{\gamma}}},
\end{equation}
which can be further reduced into the form of [27, Sec. 7.8]
\begin{equation}
\label{eq_Pd_nocop}
\overline{P}_{D} = \frac{2 \hspace{0.5 mm} e^{\frac{1}{\overline{\gamma}}}}{\Gamma(M)} \left( \frac{\lambda}{2 \hspace{0.5 mm} \overline{\gamma}} \right)^{\frac{M}{2}} K_{M} \left(\sqrt{\frac{2 \hspace{0.5 mm} \lambda}{\overline{\gamma}}}\right),
\end{equation}
where $K_{M}(.)$ is the $M^{th}$ order modified bessel function of the second kind.


\vspace{.1in}
\subsubsection{Cooperative Scheme}
A cooperative CR network consists of $N$ SUs, each senses the PU signal and reports its decision to an FC. The FC employs an \textit{n-out-of-N} fusion rule to take a final global decision. We let $l$ be the test statistic denoting the number of votes for the presence of a PU. Hence, the conditional pdfs follow a \textit{binomial distribution} \cite{5} where $P(l | \mathcal{H}_{o}) = \binom{N}{l} \hspace{1.5 mm} P_{F}^{l} \hspace{1.5 mm} (1-P_{F})^{N-l}$, and $P(l | \mathcal{H}_{1}) = \binom{N}{l} \hspace{1.5 mm} \overline{P}_{D}^{l}\hspace{1.5 mm} (1-\overline{P_{D}})^{N-l},$ where $P_{F}$ is the local false alarm probability, and $\overline{P}_{D}$ is the local detection probability averaged over the pdf of the SNR.
Based on the fusion rule mentioned above, the global false alarm and detection probabilities $P_{F_{G}}$ and $P_{D_{G}}$ are
\[P_{F,{G}} = \sum_{l=n}^{N} \binom{N}{l} P_{F}^{l} \left(1-P_{F}\right)^{N-l},\]
\begin{equation}
\label{eqn_example_det}
P_{D,{G}} = \sum_{l=n}^{N} \binom{N}{l}  \overline{P}_{D}^{l}  \left(1-\overline{P}_{D}\right)^{N-l}.
\end{equation}

\subsubsection{Single user spectrum sensing using a reconfigurable antenna}
In the proposed scheme, we assume a single SU that employs a reconfigurable antenna to sense the PU signal. Establishing the exact mathematical models for the relation between an antenna mode and the corresponding channel realization can be a daunting task. We postulate that reconfigurable antennas have an arbitrary number of possible configurations/modes (i.e., radiation patterns), and that the corresponding induced wireless channels are independent from one another (all possible radiation patterns are spatially uncorrelated). For a reconfigurable antenna with $Q$ radiation modes, we assume that $E_{i}(\Omega)$ and $E_{j}(\Omega)$ are the 3D radiation patterns corresponding to modes $i$ and $j$ respectively, and $\Omega$ is the solid angle describing the azimuth and elevation planes. Note that the solid angle ranges from 0 to $4 \pi$ steradian. The spatial correlation coefficient between the two radiation patterns is given by \cite{20}   
\[\rho_{i,j} = \frac{\int_{4 \pi} E_{i}(\Omega) E_{j}^{*}(\Omega) d \Omega}{\sqrt{\int_{4 \pi} |E_{i}(\Omega)|^{2} d \Omega \int_{4 \pi} |E_{j}(\Omega)|^{2} d \Omega}}.\]
A reconfigurable antenna is designed such that all radiation patterns are orthogonal, i.e. $\rho_{i,j} \approx 0$, $\forall i,j \in \{1,2,3,\ldots, Q\}$. For a rich scattering environment, the equivalent channel realizations encountered by different antenna states are i.i.d (independent and identically distributed) and follow a Rayleigh distribution. 
 Various designs for antennas with pattern diversity already exist \cite{16}--\cite{19}. The application of reconfigurable antennas with orthogonal patterns for MIMO systems was investigated in \cite{21}. Moreover, in \cite{22} and \cite{28}, blind interference alignment was proposed based on reconfigurable antennas with independent channels for each state. In \cite{29}, independent channel realizations were also exploited while studying the benefits of applying reconfigurable antennas in the MIMO Z interference channel. The impact of independent channel realizations perceived for different states result in a diversity gain that is similar to the spatial diversity gain attained in multiple antenna systems \cite{30}. A conceptual model for the reconfigurable antenna that resembles an antenna selection scheme is adopted in this paper. The analyses we present herein are abstract in the sense that they do not consider a specific antenna design. Fig. 1 depicts the SU receiver employing a reconfigurable antenna with $Q$ available antenna modes.

In a slow fading channel, reconfigurable antennas with $Q$ modes can offer $Q$ different channel realizations. Thus, the $i^{t h}$ sample received at the SU receiver is a binary hypothesis given by
\begin{equation}
\label{eqn_example}
   r_{i} = \left\{
     \begin{array}{lr}
       n_{i} \sim \mathcal{CN}(0, 1), & \ \mathcal{H}_{o} \\
       \sqrt{\gamma_{j}} \hspace{1 mm} S_{i} + n_{i} \sim  \mathcal{CN}(0, 1 + \gamma_{j}), & \ \mathcal{H}_{1}
     \end{array}
   \right.
\end{equation}
where $\gamma_{j} \in \{\gamma_{1}, \gamma_{2}, \cdots, \gamma_{Q}\}$ is the channel realization observed by the $i^{th}$ sample. The set of $Q$ channel gains are independent identically distributed (i.i.d.) Rayleigh random variables. It is assumed that the antenna states are switched $Q$ times within the sensing period such that channel realization $j$ is observed by $l_{j}$ samples where $\sum_{j=1}^Q l_j =M$. We designate this scheme as \textit{state switching spectrum sensing}. As an alternative, if the CSI is available at the receiver, the SU could possibly select the strongest channel for the entire sensing interval, and we call this scheme {\it state selection spectrum sensing}. Generally, the test statistic resulting at the output of the energy detector when the PU is active can be written as
\[Y = \sum_{j=1}^{L} (1+\gamma_{j}) \hspace{0.5 mm} x_{j},\]
where $L$ is the number of antenna states involved in sensing ($L \leq Q$), $\gamma_{j}$ is one of $Q$ independent channel realizations $\{\gamma_{1}, \gamma_{2}, \ldots, \gamma_{Q}\}$ assigned to the $l_{j}$ samples, and $x_{j}$ is a chi-square distributed random variable with $2 l_{j}$ degrees of freedom (the sum of $l_{j}$ normally distributed random variables). For state selection, $L$ = 1 and $l_{1} = M$ as only the highest channel gain is selected. For state switching, $L \leq Q$ and $ \sum_{j=1}^{L} l_{j} = M$. Thus, the probability of missed detection is given by 
\begin{equation}
\label{eqn_example}
P_{md}(\gamma_{1},\ldots,\gamma_{Q}) = P\left(\sum_{j=1}^{L} (1+\gamma_{j}) \hspace{0.5 mm} x_{j} \leq \lambda | \mathcal{H}_{1}, \gamma_{1},\ldots,\gamma_{Q}\right),
\end{equation}
where the threshold $\lambda$ is adjusted such that the false alarm probability $P_{F}$ = $\alpha$ in the NP test, or adjusted to minimize the error probability in the Bayesian test. It is obvious that the probability of missed detection is the cumulative distribution function (CDF) of the linear combination of chi-square random variables.
An extremely accurate approximation for the CDF of the sum of weighted chi-square random variables was proposed in \cite{34}. Based on Eqs. (20)--(23) in \cite{34}, the probability of missed detection will be given by the minimum of two functions $H(w)$ and $G(w)$ of an auxiliary parameter $w$ as follows
\[P_{md} = \min\{H(w), G(w)\},\]
where
\[w = \frac{\lambda}{M + \sum_{j=1}^{Q} l_{j} \gamma_{j}},\]
\[G(w) = \sum_{j=1}^{2M} w \hspace{0.5 mm} \frac{1+\gamma_{j}}{\lambda} \times \frac{\Upsilon\left(\frac{\lambda}{2w(1+\gamma_{j})}, \frac{\lambda}{1+\gamma_{j}}\right)}{\Gamma(\frac{\lambda}{2w(1+\gamma_{j})})},\]
and
\[H(w) = \frac{\Upsilon \left(M, \frac{\lambda}{\sqrt[M]{\prod_{i=1}^{Q}(1+\gamma_{j})^{l_{j}}}}\right)}{\Gamma(M)}.\]
 Thus, the missed detection probability in terms of the channel realizations is given by (\ref{eqn22}) where $\Upsilon(\cdot, \cdot)$ is the lower incomplete gamma function. Eq. (\ref{eqn22}) is general for any antenna state switching pattern. For state selection, the same result still applies with $l_{k} = M$, where $k = \max_{j} \gamma_{j}$ and $l_{k'}=0, k'\neq k, k'\in\{1, 2, \cdots, Q\}$.
\begin{figure*}[!t]
\normalsize
\setcounter{mytempeqncnt}{\value{equation}}
\setcounter{equation}{10}
\begin{equation}
\label{eqn22}
P_{md}(\gamma_{1},\ldots,\gamma_{Q}) = \min\left\{\frac{\Upsilon \left(M, \frac{\lambda}{\sqrt[M]{\prod_{i=1}^{Q}(1+\gamma_{i})^{l_{i}}}}\right)}{\Gamma(M)}, \,\,\, \sum_{i=1}^{2M} w \hspace{0.5 mm} \frac{1+\gamma_{i}}{\lambda} \times \frac{\Upsilon\left(\frac{\lambda}{2w(1+\gamma_{i})}, \frac{\lambda}{1+\gamma_{i}}\right)}{\Gamma\left(\frac{\lambda}{2w(1+\gamma_{i})}\right)}\right\}.
\end{equation}
\setcounter{equation}{\value{mytempeqncnt}+1}
\hrulefill
\vspace*{4pt}
\end{figure*}

\subsection{Equivalence of NP and Bayesian Optimization to Diversity Order Maximization}
The only design parameters in the spectrum sensing problem are the detection thresholds. Usually, the thresholds are selected such that the detection performance is optimized in terms of either the NP or Bayesian criteria. Obtaining the optimal detection threshold is essential for calculating the diversity order achieved by the SU receiver. However, the problem of obtaining the detection thresholds that maximize the detection or minimize the error probabilities is not always mathematically tractable, especially in the cooperative scheme \cite{7}. In this subsection, we formulate an equivalent problem for obtaining these optimal thresholds and we show that maximizing (minimizing) a performance metric $P_{*}$ is equivalent to maximizing the diversity order $d_{*}$ at assymptotically high SNR. Thus, as an alternative approach, one can obtain closed-form expressions for the diversity order $d_{*}$ in terms of the detection thresholds and get the thresholds that maximize $d_{*}$ instead of maximizing (minimizing) $P_{*}$, which is usually a mathematically tractable problem. This is formulated in the following two lemmas.\\

{\bf Lemma 1:}
{\it Based on the NP criterion, maximizing the high SNR asymptotic probability of detection under a false alarm probability constraint is equivalent to maximizing the diversity order of the detection probability.}\\

{\it proof} See Appendix A. 

{\bf Lemma 2:}
{\it Based on the Bayes detection criterion, minimizing the high SNR asymptotic probability of error is equivalent to maximizing the diversity order of the error probability.}\\
{\it Proof} See Appendix B. 

In the next section, we utilize these equivalent problems to compare the performance of the cooperative and non-cooperative schemes.

\section{To Cooperate or Not to Cooperate}
Although cooperation is widely adopted as a means of improving the performance of spectrum sensing via diversity gain, it can actually be shown that cooperative spectrum sensing does not outperform the non-cooperative scheme for the whole SNR range. Deciding whether to cooperate or not to cooperate should then depend on the operating average SNR. Specifically, for a fixed total energy constraint, the non-cooperative scheme offers a better detection performance at low SNR. This is because, at low SNR, the impact of SNR loss in the cooperative scheme due to hard decisions taken locally at each SU is higher than the gain offered by cooperation \footnote{No SNR degradation would be encountered if SUs send soft decisions to the FC. However, this is not practically feasible as the reporting channel is usually limited \cite{5}.}. On the other hand, a large diversity gain is observed at high SNR making cooperation favorable. Therefore, cooperation would not be beneficial at low SNR ranges where it is required to improve the detection performance. In addition to that, the knowledge of the number of cooperating users at each SU is essential to achieve full diversity order. Thus, even at high SNR, cooperative schemes may fail to capture full diversity gain if global network information are not provided to local SUs. In the following two subsections, we compare the two schemes and evaluate their performance in terms of diversity and coding gains, both for NP and Bayes tests.

\subsection{Non-cooperative scheme analysis}
Considering the NP test, the asymptotic expansion of $K_{M}(x)$, which appears in the $P_d$ expression in (\ref{eq_Pd_nocop}), as $x$ $\mapsto$ 0 is given by [14]
\begin{align}
K_{M}(x)&\asymp x^{-M}\Big(2^{M-1} \Gamma(M) - \frac{2^{M-3} \Gamma(M) x^{2}}{M-1} \nonumber \\ &+ \frac{2^{M-6} \Gamma(M) x^{4}}{(M-1)(M-2)} +\ldots\Big). \nonumber
\end{align}
Note that $\sqrt{\frac{2 \lambda}{\overline{\gamma}}}$ $\mapsto$ 0 and $e^{\frac{1}{\gamma}}$ $\mapsto$ 1 as $\overline{\gamma}$ $\mapsto$ $\infty$. The asymptotic expansion of the detection probability is consequently given by
\[\overline{P}_{D} \asymp 1 - \frac{\lambda}{2 \hspace{0.5 mm} \overline{\gamma} \hspace{0.5 mm} (M-1)} + \frac{\lambda^2}{8 \hspace{0.5 mm} \overline{\gamma}^2 \hspace{0.5 mm} (M-1) (M-2)} + \ldots.\]
Thus, at large average SNR, the first two terms dominate and $\overline{P}_{D} = 1 - \frac{\lambda}{2 \hspace{0.5 mm} \overline{\gamma} \hspace{0.5 mm} (M-1)} + \mathcal{O}(\overline{\gamma}^{-2})$. Hence, the average missed detection probability is $\overline{P}_{md} = 1- \overline{P}_{D} \approx \frac{\lambda}{2 \hspace{0.5 mm} \overline{\gamma} \hspace{0.5 mm} (M-1)}$. As defined in Section II, the diversity order $d$ and coding gain $A$ are, respectively, given by
\[ d_{md} = -\lim_{\overline{\gamma}\to \infty} \frac{\log \left(\frac{\lambda}{2 \hspace{0.5 mm} \overline{\gamma} \hspace{0.5 mm} (M-1)}\right)}{\log \overline{\gamma}} = 1, \]
and
\begin{equation}
\label{cods}
A = \frac{M-1}{\lambda}.
\end{equation}
Eq. (\ref{cods}) shows the diversity order and coding gain in terms of the threshold $\lambda$. It is clear that for the non-cooperative NP test, any choice of the local threshold does not affect the diversity order and the optimal threshold is selected such that it satisfies the constraint on $P_{F}$. The coding gain, on the other hand, depends on the number of samples involved in energy detection as well as the local threshold $\lambda$. The more samples involved in detection, the higher coding gain is achieved. On the other hand, large thresholds corresponding to strict false alarm constraints result in small coding gains. Note that for an $\alpha$-level NP test, the local threshold is decided by the value of $\alpha$ when setting $P_{F}$ = $\alpha$.

Now considering the Bayes optimization problem, the optimal threshold is given by the following lemma.
{\bf Lemma 3:}
{\it The optimal threshold that minimizes the average probability of error in non-cooperative spectrum sensing is given by}
\[\lambda_{opt} = \mu^{\frac{1}{M-1}}\exp\left(-\mathcal{W}_{-1}\left(\frac{-\mu^{\frac{1}{M-1}}}{2(M-1)}\right)\right),\]
at high SNR, where $\mu = \frac{P(\mathcal{H}_{1})}{P(\mathcal{H}_{o})} \times \frac{2^{M-2} \Gamma(M-1)}{\overline{\gamma}}$ and $\mathcal{W}_{-1}(.)$ is the \textit{Lambert W} function \cite{32}.\\
{\it Proof} See Appendix C. 

In order to investigate the impact of the threshold on the diversity order, we calculate the diversity order achieved with a non-optimal threshold in the following Lemma:
{\bf Lemma 4:}
{ \it For conventional spectrum sensing with a detection threshold of $\lambda = \theta\lambda_{opt}$ where $\theta \in \mathbb{R}$ and $\lambda_{opt}$ is the optimal Bayes threshold given by Lemma 3, the achieved diversity order for the Bayes optimization problem is $d_{e} = \min\{\theta, 1\}$. The corresponding false alarm and missed detection probabilities are given by Eq. (\ref{eqn_dbl_x})}.\\
{\it Proof} See Appendix D.

\begin{figure*}[!t]
\normalsize
\setcounter{mytempeqncnt}{\value{equation}}
\setcounter{equation}{12}
\[
P_{F} \asymp \frac{1}{\Gamma(M)}\left(\theta(M-1) \log\left(\frac{M-1}{\Gamma(M-1)^{\frac{1}{M-1}}}\overline{\gamma}^{\frac{1}{M-1}}\right)\right)^{M-1} \left(\frac{\Gamma(M-1)^{\frac{1}{M-1}}}{(M-1) \overline{\gamma}^{\frac{1}{M-1}}}\right)^{\theta(M-1)},
\]
and
\begin{equation}
\label{eqn_dbl_x}
P_{md} \asymp \frac{\theta}{\overline{\gamma}}\log\bigg(\frac{M-1}{\Gamma(M-1)^{\frac{1}{M-1}}} \overline{\gamma}^{\frac{1}{M-1}}\bigg).
\end{equation}
\setcounter{equation}{\value{mytempeqncnt}+1}
\hrulefill
\vspace*{4pt}
\end{figure*}

As stated in Lemma 4, for any threshold with $\theta >$ 1 (or equivalently $\lambda > \lambda_{opt}$), the maximum diversity order is achieved. However, given the expression of $P_{md}$ in Eq. (\ref{eqn_dbl_x}), the coding gain is $A_{md} = \frac{1}{\theta}$ if $\theta \geq 1$, and $A_{F} = \left(\frac{1}{\theta}\right)^{\frac{1}{\theta}}$ if $\theta \leq 1$. Thus, it is clear that the \textit{coding gain} decreases with the increase of $\theta$. Thus, the optimum Bayesian threshold corresponds to the \textit{minimum $\lambda$ that achieves the maximum diversity order $d_{e_{max}} = d_{md}$}. Because $d_{F}$ is an increasing function of $\theta$, we can obtain the optimum threshold by equating $d_{F}$ to $d_{md}$ instead of minimizing $P_{e}$, which is not mathematically tractable. The behavior of the achieved diversity order versus the factor $\theta$, that we denote as the \textit{drift factor}, is depicted in Fig. 2. It is shown that the optimal threshold corresponding to $\theta = 1$ represents the intersection of $d_{F}$ and $d_{md}$. This implies the following proposition.
\begin{figure}[!t]
\centering
\includegraphics[width=3.5 in]{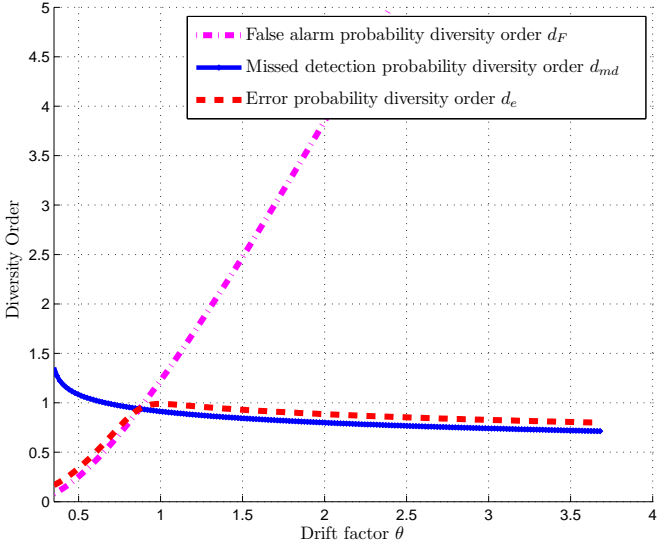}
\caption{Diversity orders ($d_{F}, d_{md}$ and $d_{e}$) versus the drift factor $\theta$ for the conventional spectrum sensing scheme.}
\label{fig_sim}
\end{figure}

\textbf{Proposition 1.} The optimal Bayesian threshold can be obtained by solving the transcendental equation
\[d_{F}(\lambda) = d_{md}(\lambda).\]

\subsection{Cooperative scheme analysis: the good, the bad, and the ugly}
In cooperative sensing, local thresholds are employed by individual SU receivers to take local hard decisions, while a global threshold (an integer number) is used by the fusion center to take the final decision. In this subsection, we relate the local and global thresholds, $\lambda$ and $n$, to the coding gain and diversity order. Next, we select the thresholds so that the global false alarm probability $P_{F,G}$ = $\alpha$ and the diversity order is maximized, which corresponds to the NP test. Then, we select the thresholds that maximize the error probability diversity order, which corresponds to the Bayesian test. We characterize the performance of energy constrained CSS as being multifaceted with three basic aspects: a ``{\it good}" aspect, which is achieving diversity order of $N$ at assymptotically high SNR; a ``{\it bad}" aspect, which is the poor coding gain causing performance degradation at low SNR; and an ``{\it ugly}" aspect, which is the inability to achieve the full diversity order when the SUs do not know the number of cooperating SUs $N$. In this case, cooperation does not reach the maximum possible diversity gain in addition to having a poor coding gain, questioning its usefulness. Hereunder, we present a comprehensive study for the performance of the cooperative scheme.

Based on (\ref{eqn_example_det}), the global missed detection probability is given by
\begin{equation}
\label{eqn_example21}
P_{md, G}(n,\lambda) = \sum_{l=0}^{n-1} \binom{N}{l} \hspace{1.5 mm} \overline{P}_{md}^{N-l}(\lambda) \hspace{1.5 mm} (1-\overline{P}_{md}(\lambda))^{l}.
\end{equation}
It is obvious that $\overline{P}_{md}$ $\mapsto$ 0 as $\overline{\gamma}$ $\mapsto$ $\infty$. The last term in the series in (\ref{eqn_example21}) dominates and the asymptotic value of $P_{md,G}$ becomes
\begin{equation}
\label{EXW}
P_{md,G}(n,\lambda) \asymp \binom{N}{n-1} \hspace{1.5 mm} \bigg( \frac{\lambda}{2 \hspace{0.5 mm} \overline{\gamma} \hspace{0.5 mm} (M-1)}\bigg)^{N-n+1}. 
\end{equation}
Thus, by rearranging (\ref{EXW}) in the form of $(A \overline{\gamma})^{-d}$, the diversity order $d_{md}$ and coding gain $A_{md}$ in terms of the local and global thresholds are given by
\[d_{md, G} = N - n + 1, \]
\[A_{md, G} \propto \binom{N}{n-1}^{\frac{-1}{N-n+1}} \hspace{1.25 mm} \frac{M-1}{\lambda}.\]
Clearly, the global threshold that maximizes the diversity order is $n$ = 1, which is known as the OR rule \cite{5}. Hence, if only one SU votes for the presence of a primary user, the fusion center adopts its decision. The local threshold $\lambda$ is chosen such that $P_{F, G}$ = $\alpha$.

Based on the above analysis, it can be concluded that cooperative spectrum sensing with $N$ SU receivers can offer a diversity order of $N$. The larger $N$ is, the higher the diversity order is, but the more information is lost due to hard decisions taken locally at each SU. This is demonstrated by the fact that the coding gain $A_{md, G} \propto M$ at $n = 1$, which is as low as $\frac{1}{N}$ of the total number of samples ($NM$) involved in detection, but the diversity gain will be maximized and $d_{md, G} = N$. In the low SNR region, information loss due to poor coding gain is more critical and we do not benefit from multiuser diversity. Thus, for a fixed total energy constraint, it is better not to cooperate when the SNR is low as assigning the total energy to a single SU leads to a better detection performance.

To demonstrate the tradeoff between coding and diversity gains, we compare a cooperative network with $N$ SU terminals and $M$ samples per terminal with a non-cooperative network with a single SU and $NM$ samples. Note that the total sensed energy is constant in both cases to ensure a fair comparison. Let the local thresholds in the multiple and single-user cases be $\lambda_{N, M}$ and $\lambda_{1,NM}$, respectively. Based on the above results, the coding gain would be $\frac{M-1}{\lambda_{N, M}}$ in the cooperative scheme and $\frac{NM-1}{\lambda_{1, NM}}$ in the non-cooperative scheme. Thus, the coding gain of the non-cooperative scheme is boosted by a factor of $N$. This factor is reduced as $\lambda_{N, M}$ and $\lambda_{1, NM}$ are not generally equal.

\begin{figure}[!t]
\centering
\includegraphics[width=3.5 in]{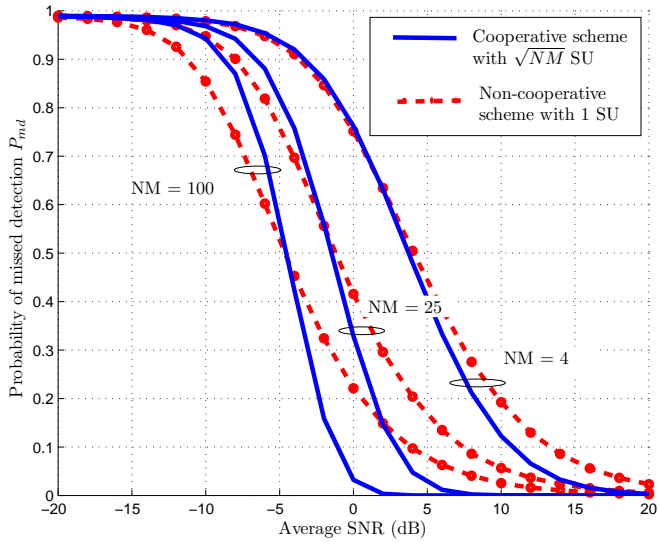}
\caption{To cooperate or not to cooperate tradeoff.}
\label{fig_sim}
\end{figure}

Fig. 3 depicts the tradeoff under study. Simulations were carried out for cooperative and non-cooperative schemes and the missed detection probability is plotted versus the average SNR. The $NM$ product is fixed for both schemes and is set to 4, 25 and 100. This product represents the total energy constraint involved in detection. For each value of $NM$, the cooperative scheme employs $\sqrt{NM}$ SU terminals and $\sqrt{NM}$ samples per terminal \footnote{Any combination of the number of SU terminals and the number of samples that keeps the $NM$ product constant can be used in the analysis.}. On the other hand, the non-cooperative scheme employs 1 SU using $NM$ samples. By applying the NP test and setting $\alpha$ = 0.01, it is found that at $NM$ = 100, the non-cooperative scheme outperforms the cooperative scheme by 3 dB at low SNR. Thus, it is better not to cooperate if the operating SNR is less than $-5$ dB, which is the SNR value corresponding to the intersection of the $P_{md}$ curves for both schemes. The SNR gain is reduced in the $NM$ = 25 scenario and nearly vanishes when $NM$ = 4. On the other hand, the cooperative scheme offers large gains in the high SNR region. For instance, at $P_{md}$ = 0.03 and $NM$ = 100, cooperation outperforms non-cooperative sensing by an SNR gain of 7 dB due to the multiuser diversity. The larger $N$ is, the more gain one gets at high SNR, but at the expense of the coding gain for a fixed energy constraint.

For the Bayesian optimization problem, we obtain the global false alarm probability by taking the dominant term of the binomial expansion in (\ref{eqn_example_det})
\[P_{F, G}(n,\lambda) \asymp \binom{N}{n} \hspace{1.5 mm} \left(\frac{\Gamma(M,\frac{\lambda}{2})}{\Gamma(M)}\right)^{n},\]
Based on the series expansion $\frac{\Gamma(M,\frac{\lambda}{2})}{\Gamma(M)} = \sum_{i=0}^{M-1} \frac{\lambda^{i}}{2^{i}\Gamma(i+1)}e^{\frac{-\lambda}{2}}$ \cite{6}, we can approximate the false alarm probability as
\begin{equation}
\label{eqn_example22}
P_{F, G}(n,\lambda) \approx \binom{N}{n} \hspace{1.5 mm} \left(\frac{\lambda^{M-1}}{2^{M-1}\Gamma(M)}\right)^{n} e^{\frac{-\lambda}{2}n}.
\end{equation}
We substitute $\lambda$ in (\ref{eqn_example21}) and (\ref{eqn_example22}) with the locally optimal threshold multiplied by the factor $\theta$. Our objective is to obtain the value of $\theta$ that maximizes the diversity order of the global error probability. The global false alarm and detection probabilities in terms of $\theta$ are given in (\ref{eqn_dbl_x35}).

\begin{figure*}[!t]
\normalsize
\setcounter{mytempeqncnt}{\value{equation}}
\setcounter{equation}{16}
\[
P_{F, G} \asymp  \binom{N}{n} \hspace{1.5 mm} \left(\frac{\left(2(M-1) \theta \log\left(\frac{M-1}{\Gamma(M-1)^{\frac{1}{M-1}}} \overline{\gamma}^{\frac{1}{M-1}}\right)\right)^{M-1}}{2^{M-1}\Gamma(M)}\right)^{n} \left(\frac{\Gamma(M-1)^{\frac{1}{M-1}}}{M-1}\right)^{\theta n (M-1)}  \frac{1}{\overline{\gamma}^{\theta n}},
\]
\begin{equation}
\label{eqn_dbl_x35}
P_{md, G} \asymp \binom{N}{n-1} \hspace{1.5 mm} \left( \frac{ \theta \log\bigg(\frac{M-1}{\Gamma(M-1)^{\frac{1}{M-1}}} \overline{\gamma}^{\frac{1}{M-1}}\bigg)}{\overline{\gamma}}\right)^{N-n+1}.
\end{equation}
\setcounter{equation}{\value{mytempeqncnt}+1}
\hrulefill
\vspace*{4pt}
\end{figure*}
From (\ref{eqn_dbl_x35}), it is obvious that $d_{md,G} = N-n + 1$, while $d_{F,G} = n \theta$. Thus, the diversity order of the error probability is
\[d_{e,G} = \min\{N-n + 1, n \theta\}.\]
We investigate the achievable diversity order in two different scenarios as follows:
\begin{itemize}
\item {\bf The number of cooperating users $N$ is unknown at SU receivers:} In this case, we aim at selecting the global threshold $n$ and the local threshold $\theta \lambda_{opt}$, such that $\theta$ is not a function of $N$. The optimal thresholds are obtained based on the following optimization problem
\[\max_{n, \theta} \,\,\, \min\{n\theta, N-n+1\}\]
\[\mbox{s.t.} \,\,\, n\theta =  N-n+1.\]
Because the number of SUs is unknown at each SU, we select a locally optimal threshold for each SU by setting $\theta = 1$. Combining this fact with {\it Proposition 1}, we obtain the optimal global threshold by solving the equation $n = N-n+1$, which yields a global threshold of $n = \lfloor \frac{N+1}{2}\rfloor$~\footnote{Throughout this paper, the operator $\lfloor .\rfloor$  is the flooring operator, while $\lceil . \rceil$ is the ceiling operator.}. Thus, the corresponding diversity order is
\[d_{e} = \min\left\{\lfloor \frac{N+1}{2}\rfloor, \lceil \frac{ N+1}{2} \rceil\right\} = \lfloor \frac{N+1}{2}\rfloor.\]
Thus, the ``ugly" face of CSS appears when global information are not provided to local SUs. Note that for $N$ = 2, cooperation without global knowledge of $N$ yields no diversity gain at all.
\item {\bf The number of cooperating users $N$ is known at SU receivers:} It is obvious that $d_{md,G}$ is maximized by setting $n$ = 1. Applying {\it Proposition 1}, the optimal value of $\theta$ is $N$. The corresponding diversity order $d_{e,G} = N$, thus the full diversity order is achieved in this case.

\end{itemize}
It is worth mentioning that global knowledge of $N$ is also needed in the NP test. However, the lack of knowledge of $N$ in the NP problem has no effect on the diversity order. Instead, it turns the problem into a {\it discrete hypothesis detection problem} \cite{31}, where only discrete values of $P_{F,G} = \alpha$ are realizable. As mentioned earlier, tolerating a larger $\alpha$ comes at the expense of the coding gain and not the diversity order.

To sum up, whether to cooperate or not to cooperate depends on several factors. If the operating SNR is low, it is better not to cooperate as the coding gain is severely degraded in the cooperative systems impacting performance at low SNR. Moreover, if the number of SUs is not known, we can not achieve the full diversity order in the Bayesian test. For small number of cooperating users (e.g., $N$ = 2), the system will not offer significant diversity gain and cooperation may not be worth it. Stemming from this analysis, we study the performance of the proposed single reconfigurable antenna schemes in the next section. Such schemes are capable of overcoming all the drawbacks of cooperation and achieving the full diversity and coding gains thus offering a superior performance compared to the conventional schemes for the entire SNR range.

\section{Spectrum Sensing Via Reconfigurable Antennas}
As stated earlier, reconfigurable antennas can artificially induce fluctuations in the slow fading channel. This would create temporal diversity for a single SU network, which can offer a gain similar to the spatial diversity gain in the cooperative scheme. We investigate two basic schemes for spectrum sensing using a reconfigurable antenna: a \textit{state switching} scheme (when the CSI is unknown) and a \textit{state selection} scheme (when the CSI is available). Based on the signal model presented in Section II, we derive the optimal test statistic for spectrum sensing with an arbitrary selection of antenna modes over time, where each mode $j$ is selected for $l_{j}$ sensing samples.

{\bf Lemma 5:}
{\it For spectrum sensing using reconfigurable antennas with arbitrary antenna state selection over time, let $Z_{j} = \sum_{i=l_{j-1}+1}^{l_{j-1}+l_{j}} |r_{i}|^{2}, j \in \{1, 2, \cdots, Q\}, l_{o} = 0,$ $L$ is the number of antenna states invoked within the sensing period ($L \leq Q$), and $\eta$ is an arbitrary detection threshold. The \textit{Likelihood Ratio Test} (LRT) reduces to}
\[ \sum_{j=1}^{L} \frac{\gamma_{j}}{1+\gamma_{j}} Z_{j} \mathop{\gtreqless}_{\mathcal{H}_{0}}^{\mathcal{H}_{1}} \eta \] \\
{\it proof} See Appendix E. 

Note that the LRT described in Lemma 5 requires the knowledge of the channel realizations corresponding to different antenna states, and involves a test statistic that is calculated via \textit{weighted energy detection} rather than simple energy detection. If the CSI is not available at the SU (i.e., the set of channel realizations $\{\gamma_{1}, \gamma_{2}, \cdots, \gamma_{Q}\}$ is unknown), the test in Lemma 5 denotes a \textit{hypothesis detection problem with unknown parameters} \cite{31}. Because the test statistic depends on the unknown parameters, no Uniformly Most Powerful (UMP) test exists, and we adopt a suboptimal test that involves simple energy detection without assigning weights to energy samples. In the state switching scheme, we blindly select an arbitrary number of channels over the sensing period such that $L \leq Q$ and $\sum_{j=1}^{L} l_{j} = M$. On the other hand, if the CSI is available at the SU, we adopt the state selection scheme instead, where the strongest channel realization is selected for the entire sensing period (i.e. $L = 1, l_{k} = M,$ and $k = \max_{j} \gamma_{j}$).

\subsection{Optimal sensing based on NP Criterion}

\subsubsection{Spectrum Sensing via State Switching}
The missed detection probability for an arbitrary antenna mode switching pattern is given by (\ref{eqn22}). Given that $\Upsilon(M, x)$ $\asymp$ $\frac{x^M}{M}$ as $\overline{\gamma}$ $\to$ $\infty$ \cite{27}, the asymptotic values of $H(w)$ and $G(w)$ are $\frac{\lambda^M}{\Gamma(M+1)\prod_{j=1}^{Q} (1+\gamma_{j})^{l_{j}}}$ and $\sum_{j=1}^{2M} w \hspace{0.5 mm} \frac{1+\gamma_{j}}{\lambda}$, respectively, which implies that $\min\{G(w), H(w)\} = H(w)$ at high SNR. Thus, one can calculate the diversity order based on $P_{md}$ = $H(w)$. The asymptotic missed detection probability will then be given by
\begin{equation}
\label{eqpo}
P_{md}(\gamma_{1},\ldots,\gamma_{Q}) \asymp \frac{\lambda^M}{\Gamma(M+1)\prod_{j=1}^{Q} (1+\gamma_{j})^{l_{j}}}.
\end{equation}
By averaging the missed detection probability in (\ref{eqpo}) over the pdf of $Q$ independent Rayleigh channel realizations we get
\[\overline{P}_{md} = \frac{\lambda^M}{\Gamma(M+1)} \int_{\gamma_{1}=0}^{\infty} \int_{\gamma_{2}=0}^{\infty} \ldots \int_{\gamma_{Q}=0}^{\infty} \frac{1}{\prod_{j=1}^{Q} (1+\gamma_{j})^{l_{j}}} \times \]
\[\frac{1}{\overline{\gamma}^Q} e^{\frac{-\sum_{j=1}^{Q} \gamma_{j}}{\overline{\gamma}}} d \gamma_{1} d \gamma_{2} \ldots d \gamma_{Q},\]
which can be reduced to
\begin{equation}
\label{eqpox}
\overline{P}_{md} = \frac{\lambda^M}{\Gamma(M+1)} \prod_{j=1}^{Q} \int_{\gamma_{j}=0}^{\infty} \frac{1}{ (1+\gamma_{j})^{l_{j}}} \frac{1}{\overline{\gamma}} e^{\frac{- \gamma_{j}}{\overline{\gamma}}} d \gamma_{j}.
\end{equation}
It can be easily shown that the integral in (\ref{eqpox}) is given by
\[\overline{P}_{md} = \frac{\lambda^M}{\Gamma(M+1)} \prod_{j=1}^{Q} \overline{\gamma}^{-l_{j}} e^{\frac{1}{\overline{\gamma}}} \Gamma \left(1-l_{j}, \frac{1}{\overline{\gamma}}\right).\]
At large SNR, $e^{\frac{1}{\overline{\gamma}}}$ $\to$ 1 and $\Gamma(1-l_{j}, \frac{1}{\overline{\gamma}})$ $\asymp$ $\frac{\overline{\gamma}^{l_{j}-1}}{l_{j}-1}$ yielding
\begin{equation}
\label{asympeq}
\overline{P}_{md} \asymp \frac{\lambda^M}{\Gamma(M+1)} \times \frac{1}{\overline{\gamma}^{Q}\prod_{j=1}^{Q} (l_{j}-1) \hspace{1 mm}}.
\end{equation}

Optimizing the coding gain depends on the choice of the number of samples $l_{i}$ associated to an antenna realization $\gamma_{i}$. It is obvious from (\ref{asympeq}) that minimizing the missed detection probability is achieved by maximizing the quantity $\prod_{i=1}^{Q} (l_{i}-1)$. We can obtain the optimum values of the $l_{i}$'s via a simple \textit{Lagrange optimization problem} as
\[
\max \hspace{1 mm} \prod_{i=1}^{Q} (l_{i}-1)\]
\[
\mbox{s.t.} \hspace{1 mm} \sum_{i=1}^{Q} l_{i} = M.
\]
By constructing the auxiliary function $\Theta(l_{1}, l_{2}, \ldots, l_{Q}, \Lambda)$ = $\prod_{i=1}^{Q} (l_{i}-1)$ + $\Lambda$ $(\sum_{i=1}^{Q} l_{i}-M)$ (where $\Lambda$ is the lagrange multiplier) and solving for $\bigtriangledown_{(l_{1}, l_{2}, \ldots, l_{Q})}\Theta(l_{1}, l_{2}, \ldots, l_{Q}, \Lambda)$ = 0 (where $\bigtriangledown$ is the gradient operator), we obtain the optimum solution as
\[l_{1} = l_{2} = \ldots = l_{Q} = \lfloor \frac{M}{Q} \rfloor.\]
Thus, the optimum antenna switching pattern is to change the antenna radiation mode every $\lfloor \frac{M}{Q} \rfloor$ samples. Note that this result is intuitive as all channel realizations are independent and identically distributed, which means that the optimal antenna mode switching pattern is obtained when employing every mode for an equal time interval during the sensing period.

From (\ref{asympeq}), the achieved diversity order is
\[d_{md} = -\lim_{\overline{\gamma}\to \infty} \frac{\log \overline{P}_{md}}{\log \overline{\gamma}} = Q.\]
Note that if the number of samples is less than the number of antenna states, only $M$ channel realizations can be employed during the sensing period. Thus, the diversity order is generally given by
\[d_{md} = \min\{M, Q\}.\]
The threshold $\lambda$ is selected such that $P_{F} = \alpha$, where it has no impact on the diversity order. The average PU signal energy input to the energy detection is given by $ Var\left\{\sum_{j=1}^{L} \sum_{i=l_{j-1}+1}^{l_{j-1}+l_{j}} \sqrt{\gamma_{j}} S_{i}\right\} = \sum_{j=1}^{L} \sum_{i=l_{j-1}+1}^{l_{j-1}+l_{j}} \overline{\gamma} = M \overline{\gamma}$. Thus, the coding gain is proportional to the total number of samples involved in detection, and the full coding gain is achieved.

\subsubsection{Spectrum Sensing via State Selection}
In the non-cooperative scheme, knowledge of the CSI at the SU can provide neither coding nor diversity gain to the detection performance. In the proposed scheme, the CSI is utilized to \textit{select} the ``best" antenna mode (the mode with largest channel gain) rather than \textit{switch} the antenna modes over time. This resembles \textit{selection combining} in multiple antenna systems. Thus, an SNR gain is obtained that is termed as a \textit{selection gain}. The pdf of the maximum of $Q$ Rayleigh distributed channel gains is given by \cite{33}
\[
f_{\gamma_{max}}(\gamma_{max}) = \frac{Q}{\overline{\gamma}} e^{\frac{-\gamma_{max}}{\overline{\gamma}}}(1-e^{\frac{-\gamma_{max}}{\overline{\gamma}}})^{Q-1}.
\]
In order to simplify the analysis, we focus on the dominant fading density at assymptotically large $\overline{\gamma}$, which can be written as [16]
\[ f_{\gamma_{max}}(\gamma_{max}) \approx \frac{Q}{\overline{\gamma}^{Q}} e^{\frac{-\gamma_{max}}{\overline{\gamma}}} \gamma_{max}^{Q-1},\]
and the probability of missed detection as a function of the instantaneous channel gain is obtained from (\ref{eqn22}) by setting $l_{k} = M$, where $k = \max_{j} \gamma_{j}$, and $\gamma_{k}$ is the corresponding channel realization
\[P_{md} = \frac{\Upsilon \left(M, \frac{\lambda}{2(1+\gamma_{k})}\right)}{\Gamma(M)}.\]
The average missed detection probability is thus given by
\begin{equation}
\label{pp}
\overline{P}_{md} = \frac{Q}{\overline{\gamma}^{Q}} \int_{\gamma_{k} = 0}^{\infty} \frac{\Upsilon \left(M, \frac{\lambda}{2(1+\gamma_{k})}\right)}{\Gamma(M)}  e^{\frac{-\gamma_{k}}{\overline{\gamma}}} \gamma_{k}^{Q-1} d\gamma_{k}.
\end{equation}
For simplicity, assume that $1+\gamma_{k}$ $\approx$ $\gamma_{k}$. The integrands in (\ref{pp}) can be represented in terms of the Meijer-G function as
\[\overline{P}_{md} = \frac{Q}{\Gamma(M) \overline{\gamma}^{Q}} \int_{0}^{\infty} \! \! \! \! \!  \gamma_{k}^{Q-1} e^{-\frac{\gamma_{k}}{\overline{\gamma}}} \MeijerG{1}{1}{1}{2}{1}{M, \hspace{0.5 mm} 0}{\frac{\lambda}{2 \hspace{0.5 mm} \gamma_{k}}} d \gamma_{k}.\]
Using the property $\MeijerG{m}{n}{p}{q}{a_1,\ldots,a_p}{b_1,\ldots,b_q}{z}$ = $\MeijerG{n}{m}{q}{p}{1-b_1,\ldots,1-b_q}{1-a_1,\ldots,1-a_p}{z^{-1}}$, the average missed detection probability will be given by the following integral
\[\overline{P}_{md} = \frac{Q}{\Gamma(M) \overline{\gamma}^{Q}} \int_{0}^{\infty} \! \! \! \! \! \gamma_{k}^{Q-1} e^{-\frac{\gamma_{k}}{\overline{\gamma}}} \MeijerG{1}{1}{2}{1}{1-M, \hspace{0.5 mm} 1}{0}{\frac{2 \hspace{0.5 mm} \gamma_{k}}{\lambda}} d \gamma_{k}.\]
Using [27, Eq. (7.813)], the average missed detection probability is
\[\overline{P}_{md} = \frac{Q}{\Gamma(M)} \MeijerG{1}{2}{3}{1}{1-Q, 1-M, \hspace{0.5 mm} 1}{0}{\frac{2 \hspace{0.5 mm} \overline{\gamma}}{\lambda}},\]
which can be represented as
\[\overline{P}_{md} = \frac{\mathcal{K}_{1}}{\overline{\gamma}^Q} \hspace{1 mm} {}_{1}F_{2}(Q; Q+1, -M+Q+1; \frac{\lambda}{2 \hspace{0.5 mm} \overline{\gamma}})\]
\begin{equation}
\label{mixedeq}
+ \frac{\mathcal{K}_{2}}{\overline{\gamma}^M} \hspace{1 mm} {}_{1}F_{2}(M; M+1, -M+Q+1; \frac{\lambda}{2 \hspace{0.5 mm} \overline{\gamma}}),
\end{equation}
where $\mathcal{K}_{1}$ and $\mathcal{K}_{2}$ are constants, ${}_{p}F_{q}(a_{1},...,a_{p}; b_{1},...,b_{q}; z)$ is the generalized hypergeometric function, and ${}_{p}F_{q}(a_{1},...,a_{p}; b_{1},...,b_{q}; z)$ $\to$ 1 as $z$ $\to$ 0. Thus, it can be easily concluded that the diversity order of the state selection scheme will be given by
\[d = \min\{M, Q\}.\]
Note that this is the same diversity order of the state switching scheme. Thus, availability of the CSI at the SU in state selection sensing offers no diversity gain compared to state switching. Selecting the best channel state every sensing period, on the other hand, offers an SNR gain (coding gain) that we define as the \textit{selection gain}. The ratio between the average SNR in the state selection scheme relative to the state switching scheme is given by
\begin{equation}
\mbox{Selection gain} = \frac{E\{\gamma_{k}\}}{E\{\gamma\}} = H_{Q},
\end{equation}
where $H_{Q}$ is the $Q^{th}$ harmonic number defined as $H_{Q} = 1 + \frac{1}{2}+\frac{1}{3}+...+\frac{1}{Q}$ \cite{33}. For large number of antenna states, the selection gain tends to
\[\mbox{Selection gain} \approx \log(Q)-\psi(1),\]
where $\psi(.)$ is the $\textit{digamma function}$ and $-\psi(1)$ is the $\textit{Euler-Mascheroni}$ constant. Thus, the coding gain obtained from state selection grows logarithmically with the number of antenna states.

\subsection{Optimal sensing based on Bayes Criterion}
\subsubsection{Spectrum Sensing via State Switching}
The achieved diversity order in this case will be obtained according to the following lemma:\\
{\bf Lemma 6:}
{\it The achieved diversity order for the proposed scheme using a threshold of $\theta \lambda_{opt}$ is $d_{e}$ = $\min\left\{\theta \min\{M,Q\}, \min\{M,Q\}\right\}$, where $\theta \in \mathbb{R}$.}\\
{\it Proof} See Appendix F. 

As stated in Lemma 6, spectrum sensing using a reconfigurable antenna with $Q$ modes can achieve a diversity order of $Q$. This is equivalent to the diversity order of a cooperative scheme with $Q$ SUs. Even if the SU is using a suboptimal threshold of $\theta \lambda_{opt}$, the achieved diversity order is $\theta Q$ which is $Q$ times larger than the diversity order achieved by the conventional scheme that employs a threshold of $\theta \lambda_{opt}$.

\subsubsection{Spectrum Sensing via State Selection}
By observing Eq. (17), the missed detection diversity order is given by  $\min \{Q,M\}$. The same diversity analysis applied for the state switching scheme in Lemma 6 can be carried out for the state selection scheme. In fact, both schemes have the same diversity order and the same optimal threshold at high SNR. Similar to the NP problem, the state selection scheme offers an extra coding gain as the average SNR is boosted by a factor of $H_{Q}$.

\subsection{Impact of Switching Delay}
In this subsection, we quantify the impact of switching delay on the detection performance of state switching and state selection schemes. Let $D$ be the equivalent number of samples that a particular switching device needs to change from one antenna state to the other. We assume that throughout those $D$ samples, the old channel realization is perceived by the SU receiver. A new channel realization appears after $D$ samples, which means that the maximum achievable switching rate is $\frac{1}{DT_{s}}$, where $T_{s}$ is the system sampling period. \footnote{Various switching devices experience different ranges of time delay. For instance, a MEMS switch may have a switching time of $10-20$ $\mu$s \cite{17}. Other electronic switching devices, such as PIN diodes or field-effect transistors (FETs), can offer a much faster switching speed \cite{20}.}

\subsubsection{Impact on state switching scheme}
In the state switching scheme with $D$ delay samples, the achieved diversity order is
\[d = \min\{Q, \frac{M}{D}\}.\]
The SU tries to rapidly switch the antenna modes such that maximum number of channel realizations is utilized in sensing. The limited switching speed affects the achieved diversity order negatively. The number of samples $l_{j}$ assigned to a channel realization $j$ must be greater than $D$. The maximum number of channel realizations that can appear within $M$ sensing samples is thus $\frac{M}{D}$.
If $Q > \frac{M}{D}$, we can not achieve the maximum diversity order. In fact, if the sensing period is limited compared to the switching delay, the diversity gain offered by reconfigurable antennas becomes less significant. If $M = D$, the system behaves like the conventional non-cooperative scheme.

\subsubsection{Impact on state selection scheme}
If the SU requires $D$ samples to select the maximum channel realization, the achieved SNR gain is perceived for $M-D$ samples only. In this case, the selection gain tends to
\[\mbox{Selection gain} = \frac{D}{M}+\frac{M-D}{M} H_{Q}.\]
Moreover, the diversity order is also impacted as the effective sensing period that is subject to the selected channel is $M-D$ samples only. Hence, the diversity order becomes
\[d = \max\{1,\min\{M-D, Q\}\}.\]
Again, at $M=D$, the system acts in an identical way to the legacy single antenna non-cooperative scheme as all samples experience an arbitrary channel without selection. Thus the dominating diversity order is either 1, when $M=D$, or $\min\{M-D, Q\}$ otherwise.
The switching delay degrades the diversity order of the state switching scheme, and both the diversity order and selection gain of the state switching scheme. The design of the reconfigurable antenna should take into account the possible values of the sensing period. It is essential to employ high speed switching devices with switching times that are significantly smaller than the sensing period. If the switching speed is inevitably low, one has to extend the sensing period such that diversity and coding gain benefits of the reconfigurable antenna are attained. However, this will be at the expense of the system throughput.

\subsection{Performance Evaluation}

In this subsection, we evaluate the performance of the proposed schemes and compare them with the conventional cooperative and non-cooperative schemes. It is important to note that all the parameter settings used in the simulations discussed in this section are selected arbitrarily for numerical and simulation convenience. However, the analyses and explanations presented in the paper are generic and suit any practical values for the system parameters. For all curves, Monte Carlo simulations are carried out with 1,000,000 runs. In Fig. 4, we plot the error probability curves for the non-cooperative, the cooperative (with $N$ being known and unknown), the state switching as well as the state selection (with $Q$ = 15 antenna states) schemes. An overall energy constraint is imposed by fixing the total number of samples to 30. It is shown that the cooperative scheme with 15 cooperating SUs outperforms the non-cooperative scheme at high SNR as it achieves a diversity order of 15. However, cooperation performs worse at SNR values below $-5$ dB due to the poor coding gain. When the number of SUs is unknown, a diversity order of $\lfloor \frac{15+1}{2} \rfloor$ =  8 is only achieved. Thus the offered diversity gain at high SNR is generally less than that offered when $N$ is known. State switching and selection are shown to outperform cooperative and non-cooperative schemes at any SNR. For state switching, a diversity order of $\min\{15,30\}$ = 15 is achieved, which is the same diversity order of the cooperative scheme, leading both curves to have the same slope. However, the state switching scheme uses 30 samples for sensing, which maintains the same coding gain of the non-cooperative scheme. It is shown that state switching acts like a non-cooperative scheme at low SNR, and provides a diversity gain at high SNR. As for the state selection scheme, it attains the same diversity order of $\min\{15,30\}$ = 15, and in addition, offers a coding gain of $H_{15} = 1 + \frac{1}{2}+\ldots+\frac{1}{15} \approx$ 5 dB. Thus, an SNR gain of about 5 dB compared to state switching is obtained via antenna state selection. Similar simulations are carried out for the NP test with 100 samples, false alarm probability of 0.05, and 10 antenna states. Fig. 5 shows that state switching and selection act in a similar manner to that depicted by Fig. 4. Again, state selection scheme outperforms all other schemes, while state switching still offers a better performance than cooperative and non-cooperative schemes. Although achieving the selection gain requires channel estimation and appropriate reconfigurable antenna design (with large number of independent states), it is still less complex than the cooperation scenario.

Fig. 6 demonstrates the impact of switching delay on the sensing performance based on the NP test for $Q$ = 10 states. For the state switching scheme, switching delay has no impact on the coding gain. However, the diversity order is reduced when the delay is introduced. For a total number of sensing samples $M$ = 100, we study the effect of the switching delay with values $D = \{30, 50, 95, 100\}$ samples. For those delay values, the delay-free diversity order of 10 is reduced to be $d_{md} = \min\{10, \frac{100}{30}\} \approx 3$, $\min\{10, \frac{100}{50}\} = 2$, $\min\{10, \frac{100}{95}\} \approx 1$, and $\min\{10, \frac{100}{100}\} = 1$, respectively. This is demonstrated by the degradation of the slope of the solid curves in Fig. 6 as delay increases. When the delay samples are equal to the sensing samples, state switching performs like the non-cooperative scheme with legacy antenna. When a very large delay of $95$ samples is encountered, the SU does not achieve any diversity gain (it will be shown later that state selection is less sensitive to large delay scenarios). At low SNR, all curves coincide as switching delay has no impact on the coding gain. Contrarily, the diversity order of the state switching scheme is less sensitive to delay and its coding gain degrades with increasing delay. For a delay of 30 samples, the full diversity order is achieved as $\max\{1, \min\{100-30, 10\}\} = 10$. However, the selection gain drops from $H_{10}= 4.667$ dB to $\frac{7}{10} H_{10} + \frac{3}{10} = 3.711$ dB. Similarly, a delay of 50 samples degrades the coding gain but preserves the diversity order. This is depicted in Fig. 6 by the three dashed curves corresponding to delays of $D$ = 0, 30, and 50 samples. The three curves have the same slope (same diversity order) but different coding gains. When the delay becomes as large as 95 samples, the diversity order drops to $\max\{1, \min\{100-95, 10\}\} = 5$, which is reflected in Fig. 6 by a significant change in the slope of $P_{md}$. It is worth mentioning that for 95 delay samples, state switching does not achieve any diversity gain, which is not the case in state selection. Thus, state selection loses its diversity gain advantages only for significantly large switching delays, but at the expense of the CSI estimation complexity. Fig. 7 shows the impact of delay on the error probability in the Bayesian test, and it is easy to interpret the results in a similar manner. 

\begin{figure}[!t]
\centering
\includegraphics[width=3.5 in]{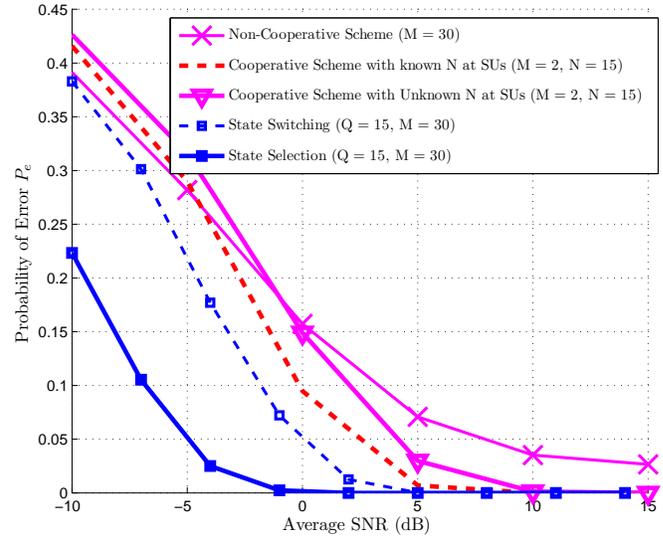}
\caption{Performance of various schemes based on the Bayesian test.}
\label{fig_sim}
\end{figure}

\begin{figure}[!t]
\centering
\includegraphics[width=3.5 in]{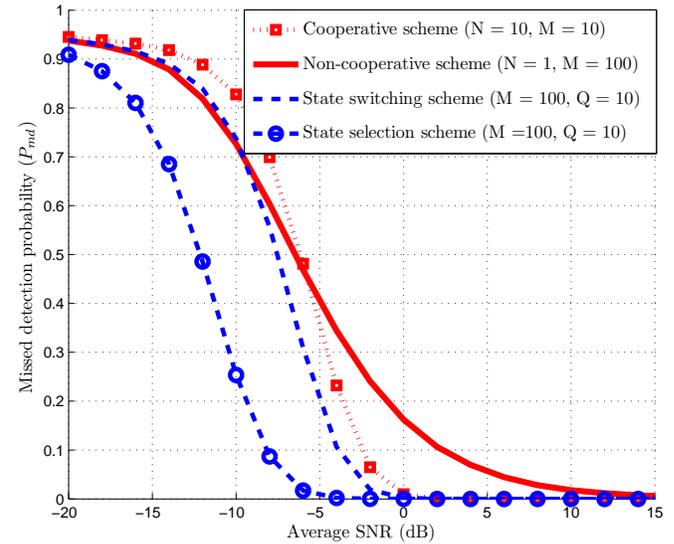}
\caption{Performance of various schemes based on NP test with $\alpha=0.05$}.
\label{fig_sim}
\end{figure}

\begin{figure}[!t]
\centering
\includegraphics[width=3.5 in]{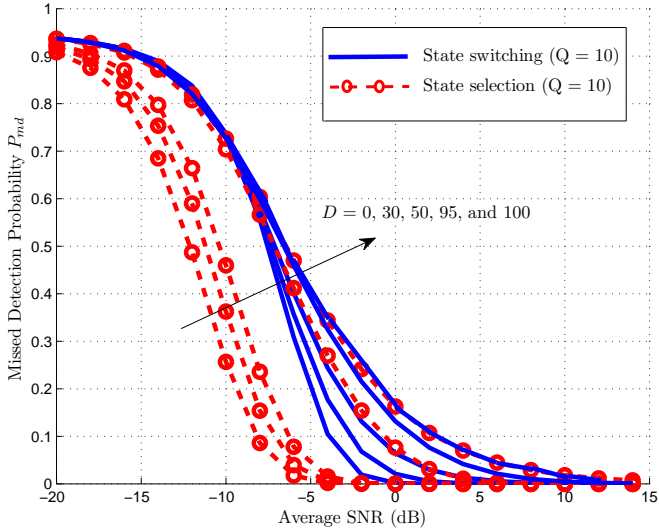}
\caption{Impact of switching delay on proposed schemes based on the NP test with M = 100 and $\alpha=0.05$.}
\label{fig_sim}
\end{figure}

\begin{figure}[!t]
\centering
\includegraphics[width=3.5 in]{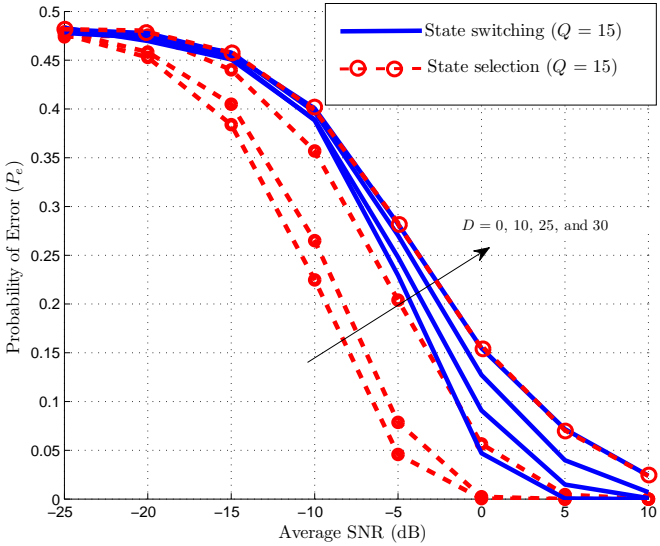}
\caption{Impact of switching delay on proposed schemes based on the Bayesian test.}
\label{fig_sim}
\end{figure}

\section{Sensing-throughput trade-off: Throughput Gain in Reconfigurable Antenna Schemes}
\begin{figure}[!t]
\centering
\includegraphics[width=3 in]{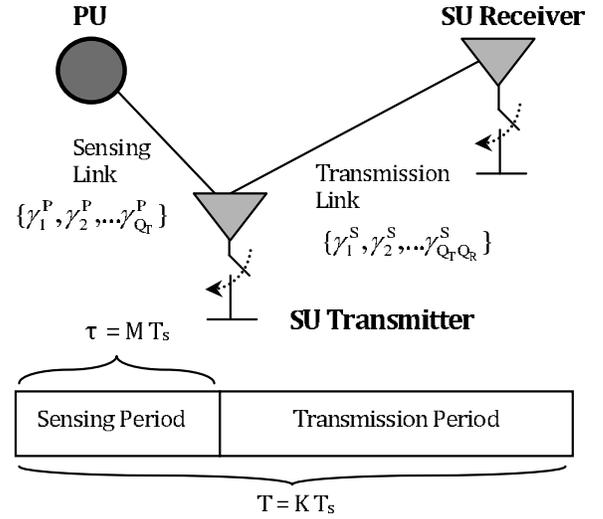}
\caption{Sensing and transmission stages in a CR system.}
\label{fig_sim}
\end{figure}
In this section, we revisit the fundamental tradeoff between sensing capability and achievable throughput of the secondary networks. We will show that there exists an optimal sensing time for which the highest throughput for the secondary network is achieved with sufficient protection for the PU. Next, we will show that by adopting state selection spectrum sensing, this optimal sensing time is reduced, thus allowing for an even higher throughput given the same PU protection constraints. Furthermore, we show that the SU transmitter and receiver can utilize reconfigurable antennas to maximize the secondary channel capacity by selecting the ``best" antenna states at both secondary parties. Thus, not only do reconfigurable antennas improve the performance in the detection phase, but they can also be utilized to enhance the channel capacity in the transmission phase as well. Finally, we investigate the effect of switching delay on the achievable capacity and quantify the possible degradation caused by such delay.

The sensing-throughput tradeoff was studied thoroughly in \cite{23}--\cite{25}. We are concerned here with the impact of reconfigurable antenna spectrum sensing on throughput given a constraint on the detection probability. In the next subsections, we compare the reconfigurable antenna state selection scheme with the conventional one. We omit CSS from our discussion for fair comparison, as the throughput achieved by CSS is divided among the cooperating users. Besides, we only consider state selection and not state switching, as the constraints on detection probability are usually given at low SNR \cite{23}, which takes away any advantage of state switching. In addition to that, it is obvious that state switching can not improve the ergodic capacity as it has no CSI. 

\subsection{Problem Formulation}
As depicted by fig. 8, we assume a frame structured secondary network consisting of an SU transmitter and an SU receiver. The frame is divided into a sensing period of length $\tau$ and a transmission period of $T-\tau$. The SU transmitter senses the PU signal for a period of $\tau$ and if the PU is absent, the SU transmitter sends data to the SU receiver in a period of $T-\tau$. For a sampling period of $T_{s}$, we have $\tau = M T_{s}$ and $T = K T_{s}$, where $M$ is the number of samples used in sensing and $K$ is the total number of samples in the frame. We assume that the SU transmitter employs a reconfigurable antenna with $Q_{T}$ states, while the SU receiver has a reconfigurable antenna with $Q_{R}$ states. The SU transmitter is engaged in two phases:
\begin{itemize}
\item {\bf Sensing phase:} where the SU transmitter senses the PU signal after applying state selection and selects the strongest channel out of the $Q_{T}$ channel realizations $\{\gamma_{1}^{P}, \gamma_{2}^{P}, \ldots, \gamma_{Q_{T}}^{P}\}$ between the SU transmitter and the PU.
\item {\bf Transmission phase:} where the SU transmitter and receiver apply state selection jointly and select the strongest channel out of $Q_{T} Q_{R}$ possible channel realizations $\{\gamma_{1}^{S}, \gamma_{2}^{S}, \ldots, \gamma_{Q_{T}Q_{R}}^{S}\}$.
\end{itemize}
Thus, the SU transmitter selects the best antenna state for sensing and then switches to the best state for transmission. We assume the availability of full CSI at the SU parties. If switching delay is considered, then $D$ samples are wasted to switch between the different modes.

\subsection{Normalized throughput maximization}
The average throughput for the secondary network as a function of the sensing period is given by \cite{23}
\begin{equation}
\label{eq}
R(\tau) = \left(1-\frac{\tau}{T}\right) \Big\{C_{o}P(\mathcal{H}_{o})(1-P_{F}(\tau))+C_{1}P(\mathcal{H}_{1})P_{md}(\tau)\Big\}.
\end{equation}
If $\gamma_{p}$ is the channel between SU receiver and the PU, and $\gamma_{s}$ is the secondary transmission channel, then $C_{o} = \log(1+\gamma_{s})$ and $C_{1} = \log(1+\frac{\gamma_{s}}{1+\gamma_{p}})$. Because $P(\mathcal{H}_{1})$ is usually less than $P(\mathcal{H}_{o})$ and $C_{1} < C_{o}$, a reasonable approximation for $R(\tau)$ is adopted in \cite{23}--\cite{24} as
\begin{equation}
\label{eq25_ed}
R(\tau) \approx C_{o}P(\mathcal{H}_{o})\left(1-\frac{\tau}{T}\right)(1-P_{F}(\tau)).
\end{equation}
From (\ref{eq25_ed}), we note that two factors affect the average secondary throughput. First, as the sensing time increases, the throughput decreases as less time is dedicated to transmission within a frame. Second, a high value for the false alarm probability degrades the throughput as it implies that we waste opportunities to access the channel. The average normalized throughput is defined as $\tilde{R}(\tau) = \frac{R(\tau)}{C_{o} P(\mathcal{H}_{o})}$, which can be expressed as  
\[\tilde{R}(\tau) = \frac{T-\tau}{T} (1-P_{F}(\tau)) .\]
 The optimal sensing time is obtained by maximizing $\tilde{R}(\tau)$ while keeping $\overline{P}_{D}(\tau)$ above a certain threshold
\[ \max \,\,\, \tilde{R}(\tau)\]
\begin{equation}
\label{eq20}
 \mbox{s.t.} \,\,\, \overline{P}_{D}(\tau) \geq p_{d}.
\end{equation}
It is easy to prove that $\tilde{R}(\tau)$ has a unique maximum by proving its unimodality. The derivative of $\tilde{R}(\tau)$ with respect to $\tau$ is given by
\begin{equation}
\label{eq21}
\frac{ \partial \tilde{R}(\tau)}{\partial \tau} = \underbrace{\frac{-1}{T}(1-P_{F}(\tau))}_{A_{1}}+\underbrace{(1-\frac{\tau}{T})\left(-\frac{d P_{F}(\tau)}{d \tau}\right)}_{A_{2}}.
\end{equation}
Notice that the term $A_{1}$ is always negative as $P_{F}(\tau)$ is always less than 1. Also, as $P_{F}(\tau)$ decreases with increasing $\tau$, then $A_{1}$ is a monotonically decreasing function of $\tau$. As for the term $A_{2}$, it is always positive because $P_{F}(\tau)$ is a monotonically decreasing function in $\tau$, which means that $-\frac{d P_{F}(\tau)}{d \tau}$ is always positive. Moreover, as $\tau < T$, then $(1-\frac{\tau}{T})$ is also positive and $A_{2}$ is positive for all $\tau$. Finally, it can be shown that $-\frac{d P_{F}(\tau)}{d \tau}$ is a monotonically decreasing function of $\tau$, thus $A_{2}$ is also monotonically decreasing in $\tau$. Now, the sum of the two monotonic functions $A_{1}$ and $A_{2}$ is positive if $|A_{2}| > |A_{1}|$ and negative otherwise. Therefore $\tilde{R}(\tau)$ is unimodal and has an extremum point at $|A_{2}| = |A_{1}|$.

It is shown in \cite{23} that the optimal solution to (\ref{eq20}) is achieved with equality constraint. Assume that for the conventional spectrum sensing scheme, the optimal number of sensing samples is $M_{opt}$. For this number of samples, the detection probability satisfies the equality constraint $\overline{P}_{D}(\tau) = p_{d}$. For state selection spectrum sensing with $Q_{T}$ antenna states, we have shown that a coding gain of $10 \log(H_{Q_{T}})$ dB is obtained. Thus, to satisfy the constraint of $\overline{P}_{D}(\tau) = p_{d}$ with state selection at low SNR, we only need $\frac{M}{H_{Q_{T}}}$ samples for sensing. If the optimal sensing time for the conventional scheme is $M_{opt}$ and the corresponding false alarm probability is $P_{F,c}$, and if the false alarm probability of the state selection scheme with $\frac{M_{opt}}{H_{QT}}$ sensing samples is $P_{F,s}$, then the normalized throughput gain is
\[\mbox{Normalized througput gain} = \frac{1-\frac{M_{opt}}{K H_{QT}}}{1-\frac{M_{opt}}{K}} \times \frac{1-P_{F,s}}{1-P_{F,c}}.\]
Note that $P_{F,s}$ is always less than $P_{F,c}$ for a constant detection probability. The reason for this is that, for a fixed threshold $\lambda$, we have $\frac{\Gamma(M_{opt}, \frac{\lambda}{2})}{\Gamma(M_{opt})} > \frac{\Gamma(\frac{M_{opt}}{H_{Q_{T}}}, \frac{\lambda}{2})}{\Gamma\left(\frac{M_{opt}}{H_{Q_{T}}}\right)}$ as the false alarm probability is a monotoically decreasing function of the number of sensing samples. In addition to that, the state selection scheme offers a diversity gain, which means that even when the sensing samples are only $\frac{M_{opt}}{H_{QT}}$, the state selection scheme still outperforms the conventional scheme with $M_{opt}$ samples at any SNR. Thus, for a fixed detection probability, the optimal threshold in the state selection scheme is greater than that used in the conventional scheme. Therefore, the false alarm probability is reduced by state selection even if the detection probability is kept constant. This means that by using reconfigurable antennas, a multifaceted throughput gain is achieved. For a fixed detection probability, the optimal sensing time is reduced allowing for longer transmission period, and the false alarm probability is reduced, which in turn, means a better utilization of the channel when the PU is absent.

\begin{figure}[!t]
\centering
\includegraphics[width=3.5 in]{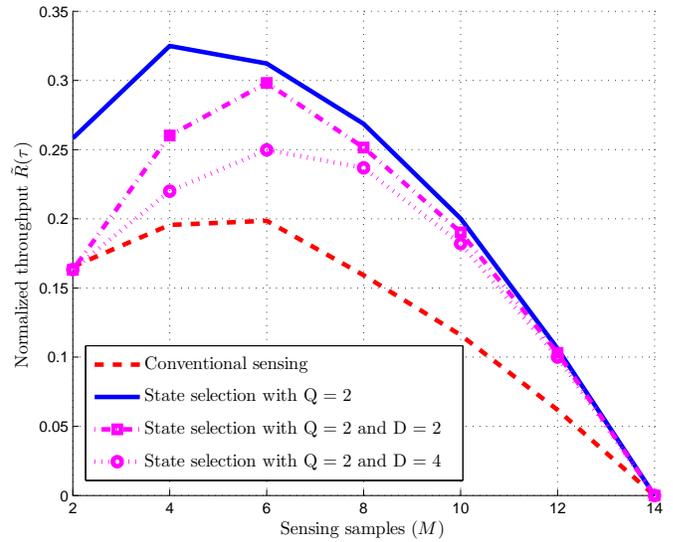}
\caption{Optimal sensing time in conventional and state selection schemes (SNR = 0 dB and $\overline{P}_{D}$ = 0.9).}
\label{fig_sim}
\end{figure}

Fig. 9 depicts the normalized throughput gain obtained by deploying state selection with $Q$ = 2. Assuming that the detection probability is set to 0.9 at an average SNR of 0 dB, the normalized throughput curves for conventional and state selection schemes are plotted versus the number of samples $M$. It is shown that the optimal sensing time for the conventional scheme is $M$= 6, which is reduced to 4 in the state selection scheme as the required number of samples to attain the same detection probability becomes $\frac{6}{H_{2}} = 4$. Besides, the false alarm improvement in the state selection scheme contributes to the total throughput gain. It can be deduced from the peak values that the maximum normalized throughput is boosted from 0.2 to 0.325 when state selection is applied. This gain degrades when switching delay is considered, which is depicted in Fig. 9 for $D$ = 2 and 4. For $D$ = 2, the maximum normalized throughput drops from 0.325 to 0.3, while a delay of $D$ = 4 results in a maximum normalized throughput of 0.25 only.

\subsection{Transmission Channel Capacity}
In the previous subsection, we demonstrated the normalized throughput gain achieved by using a reconfigurable antenna in the sensing phase. It is worth mentioning that the SU transmitter can select different antenna states for sensing and transmission to achieve diversity in PU signal detection and SU-to-SU signal transmission. The maximum achievable average throughput is approximated as
\[
R = \sup_{1\leq i \leq Q_{T}, 1\leq j \leq Q_{R}}\left(1-\frac{M}{K}\right) P_{F} P(\mathcal{H}_{o}) E\big\{\log(1+\gamma_{i,j}^{S})\big\},
\]
where $\gamma_{i,j}^{S}$ is the SU transmitter and receiver channel that corresponds to transmitter and receiver antenna states $i$ and $j$, where $1\leq i \leq Q_{T}$ and $1\leq j \leq Q_{R}$. We drop the term $(1-\frac{M}{K}) P_{F} P(\mathcal{H}_{o})$ as it depends on the selected antenna state in the sensing phase. We assume that all possible $Q_{T}Q_{R}$ channel realizations are independent and identically distributed (which matches with the conceptual model in Section II), and that the average SNR of the SU link is $\overline{\gamma}_{S}$. The average (ergodic) transmission channel capacity $E\big\{\log(1+\gamma_{i,j}^{S})\big\}$ depends on the pdf of the selected antenna state. By selecting the maximum channel out of $Q_{T}Q_{R}$ channel realizations, the pdf of $\gamma = \max_{1\leq i \leq Q_{T}, 1\leq j \leq Q_{R}} \{\gamma_{1,1}^{S}, \gamma_{1,2}^{S}, \ldots, \gamma_{1,Q_{R}}^{S}, \gamma_{2,1}^{S}, \ldots, \gamma_{Q_{T},Q_{R}}^{S} \}$ is given by \cite{33}
\[f_{\gamma}(\gamma) = \frac{Q_{T}Q_{R}}{\overline{\gamma}_{S}} e^{\frac{-\gamma}{\overline{\gamma}_{S}}} (1-e^{\frac{-\gamma}{\overline{\gamma}_{S}}})^{Q_{T}Q_{R}-1},\]
which can be rewritten using the binomial theorem as
\begin{align}
\label{ex23_eqtn}
f_{\gamma}(\gamma) = Q_{T}Q_{R} \sum_{i=0}^{Q_{T}Q_{R}-1} \binom{Q_{T}Q_{R}}{i} \frac{(-1)^{i}}{\overline{\gamma}_{S}} e^{-\frac{\gamma (i+1)}{\overline{\gamma}_{S}}}.
\end{align}
Thus, the ergodic capacity $C_{s}$ of the state selection transmission is given by averaging shannon capacity over the pdf in (\ref{ex23_eqtn})
\begin{align}
\label{ex23}
C_{s}\!=\!Q_{T}Q_{R}\!\!\!\!\! \sum_{i=0}^{Q_{T}Q_{R}-1} \!\!\!\binom{Q_{T}Q_{R}}{i} \frac{(-1)^{i}}{i+1}\!\! \int_{\gamma = 0}^{\infty}\!\!\!\! \!\log(1+\gamma) \frac{e^{-\frac{\gamma (i+1)}{\overline{\gamma}_{S}}}}{\overline{\gamma}_{S}/(i+1)} d \gamma.
\end{align}
The ergodic capacity of the conventional single antenna scheme is given by $C = e^{\frac{1}{\overline{\gamma}_{S}}} \operatorname{Ei}\bigg(\frac{1}{\overline{\gamma}_{S}}\bigg)$ \cite{35}, where $\operatorname{Ei}(x) = - \int_{-x}^{\infty}\frac{e^{-t}}{t} dt$ is the exponential integral function. Thus, the ergodic capacity of the state selection scheme is given by
\begin{equation}
\label{ex25}
C_{s} = Q_{T}Q_{R} \sum_{i=0}^{Q_{T}Q_{R}-1} \binom{Q_{T}Q_{R}}{i} \frac{(-1)^{i}}{i+1} e^{\frac{i+1}{\overline{\gamma}_{S}}} \operatorname{Ei}\bigg(\frac{i+1}{\overline{\gamma}_{S}}\bigg).
\end{equation}

\begin{figure}[!t]
\centering
\includegraphics[width=3.5 in]{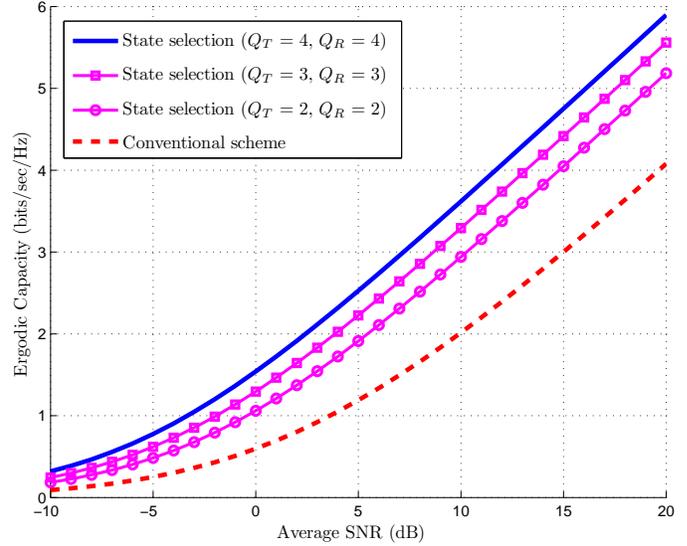}
\caption{Capacity gains for various numbers of antenna states.}
\label{fig_sim}
\end{figure}		
		
Assuming that the SU transmitter applies equal power allocation for simplicity, Fig. 10 shows the ergodic capacity gain achieved by state selection for various number of combinations of antenna states. The capacity gain becomes more significant at high SNR. For instance, at an SNR of 10 dB, the capacity of state selection with 4 antenna states is 1.75 times the conventional scheme capacity. This gain can be transformed into an SNR gain of 7.5 dB. In other words, the transmission rate of the conventional scheme at an SNR of 10 dB can be achieved by state selection at an SNR of only 2.5 dB.

For a switching delay of $D$, the SU transmits on two parallel channels: the channel utilized for sensing is still effective for the first $D$ samples of the transmission period, and the best transmission channel becomes effective for the remaining $K-M-D$ samples. The effective average capacity in this case is given by (\ref{eqn_dbl_x37}) on top of the next page. Note that when the SU transmits on the previously selected sensing channel for the first $D$ samples, it attains the same capacity of the conventional scheme. Fig. 11 demonstrates the impact of switching delay on the average capacity of state selection with 4 antenna states. When the proportion of switching delay to the total transmission time is 0.2, the capacity gain at SNR = 10 dB reduces from 1.75 to 1.625. Moreover, if the switching delay reaches half of the transmission time, the capacity gain reduces to 1.375. We infer from Fig. 11 that as long as the proportion of the switching delay to the total transmission time is less than 0.2, the SNR loss is less than 1 dB. The effect of switching delay on the achieved capacity depends on the transmission period and the switching technology. An electronic switching device should be adopted if the transmission period is comparable to the switching delay of MEMS switches.

\begin{figure}[!t]
\centering
\includegraphics[width=3.5 in]{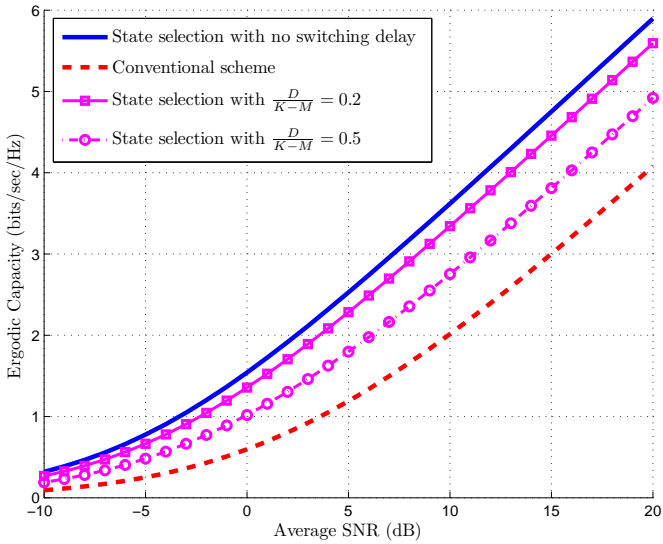}
\caption{Impact of switching delay on the average capacity ($Q_{T} = Q_{R}$ = 4).}
\label{fig_sim}
\end{figure}

\begin{figure*}[!t]
\normalsize
\setcounter{mytempeqncnt}{\value{equation}}
\setcounter{equation}{29}
\begin{equation}
\label{eqn_dbl_x37}
C_{s, D} = \frac{D}{K-M} e^{\frac{1}{\overline{\gamma}_{S}}} \operatorname{Ei}\bigg(\frac{1}{\overline{\gamma}_{S}}\bigg) +\frac{K-M-D}{K-M} Q_{T}Q_{R} \sum_{i=0}^{Q_{T}Q_{R}-1} \binom{Q_{T}Q_{R}}{i} \frac{(-1)^{i}}{i+1} e^{\frac{i+1}{\overline{\gamma}_{S}}} \operatorname{Ei}\bigg(\frac{i+1}{\overline{\gamma}_{S}}\bigg).
\end{equation}
\setcounter{equation}{\value{mytempeqncnt}}
\hrulefill
\vspace*{4pt}
\end{figure*}

\section{Conclusions}
In this paper, we discussed a tradeoff between the diversity and coding gains achieved in various spectrum sensing schemes. By obtaining the diversity and coding gains in terms of the detection thresholds, we proved that cooperative schemes are not always beneficial as hard decisions taken at local SUs cause loss of coding gain, which can be significant at low SNR. Based on this analysis, we proposed a novel spectrum sensing scheme that utilizes a reconfigurable antenna at the SU to exploit the diversity of its radiation states, achieving full diversity and coding gains without SU cooperation. The proposed scheme can outperform cooperative sensing, which involves significant overhead, at all SNR ranges. Two schemes based on reconfigurable antennas were presented: state switching and state selection. Based on a conceptual model for the reconfigurable antenna, we obtained the fundamental limits on the achievable diversity order, throughput, and transmission capacity for the proposed schemes. Furthermore, the impact of the state switching delay on the detection performance and the achievable capacity was quantified. It was shown that even with significant switching delay, detection and throughput gains are still attainable.

\appendices

\section{Proof of Lemma 1}
\renewcommand{\theequation}{\thesection.\arabic{equation}}
The NP optimization problem is formulated as
\[\max_{\lambda} \,\, \overline{P}_{d}(\lambda)\equiv \min_{\lambda} \,\, \overline{P}_{md}(\lambda)\]
\[\mbox{s.t.}\,\, P_{F} \leq \alpha,\]
where $\overline{P}_{md}(\lambda)$ is the missed detection probability as a function of the detection threshold. It follows from the definition of the diversity order in Section II that ${d}_{md} =  -\lim_{\overline{\gamma}\to \infty} \frac{\log(\overline{P}_{md}(\lambda))}{\log \overline{\gamma}}$. Note that there is a one-to-one mapping between $f(x)$ and $\log(f(x))$, and that the $\log(\cdot)$ function preserves monotonicity. Thus, maximizing $\overline{P}_{md}(\lambda)$ is equivalent to maximizing $\log(\overline{P}_{md}(\lambda))$. Dividing the objective function by the constant $\log\overline{\gamma}$ yields the equivalent problem
\[\max_{\lambda} \,\, -\frac{\log(\overline{P}_{md}(\lambda))}{\log\overline{\gamma}}\]
\begin{equation}
\label{App2_1}
\mbox{s.t.}\,\, P_{F} \leq \alpha.
\end{equation}
It is clear that as $\overline{\gamma}$ $\to$ $\infty$, the optimization problem tends to maximizing the diversity order. This concludes the proof of the lemma.

\section{Proof of Lemma 2}
\renewcommand{\theequation}{\thesection.\arabic{equation}}
The Bayesian optimization problem is equivalent to minimizing the average probability of error, viz.,
\[\min_{\lambda} \,\, \overline{P}_{e}(\lambda) = P(\mathcal{H}_{1}) \, \overline{P}_{md} + P(\mathcal{H}_{o}) \, P_{F}.\]
Recall that the receiver operating characteristics (ROC) (the plot of $P_{F}$ versus $\overline{P}_{D}$) is a strictly concave and monotonically increasing function \cite{31}, which implies the following
\begin{equation}
\label{App3_1}
\frac{d P_{F}(\lambda)}{d \overline{P}_{D}(\lambda)} > 0, \,\, \mbox{and} \,\, \frac{d P_{F}(\lambda)/d \lambda}{d \overline{P}_{D}(\lambda)/d \lambda} > 0.
\end{equation}
Because $\frac{d P_{F}(\lambda)/d \lambda}{d \overline{P}_{D}(\lambda)/d \lambda}$ is always positive, we deduce that $\frac{d P_{F}(\lambda)/d \lambda}{d \overline{P}_{md}(\lambda)/d \lambda}$ is always negative. Thus, the derivatives $d P_{F}(\lambda)/d \lambda$ and $d \overline{P}_{md}(\lambda)/d \lambda$ have opposite signs, i.e., opposite monotonic behaviors. Therefore, we conclude that the average error probability $P_{e}(\lambda) = P(\mathcal{H}_{1}) \, \overline{P}_{md} + P(\mathcal{H}_{o}) \, P_{F}$ is a unimodal function and the optimal threshold can be obtained by solving the equation
\begin{equation}
\label{App3_2}
\frac{d P_{e}(\lambda)}{d \lambda} = 0.
\end{equation}
Considering the derivative of $\log(P_{e})$ instead of $P_{e}$ yields
\[\frac{d \log(P_{e}(\lambda))}{d \lambda} = \frac{1}{P_{e}(\lambda)} \frac{d P_{e}(\lambda)}{d \lambda} = 0,\]
which is equivalent to (\ref{App3_2}), thus the Bayesian optimization problem at high SNR reduces to trying to find the threshold $\lambda^*$ such that
\begin{equation}
\label{App3_3}
\lambda^* = \max_{\lambda} d_{e},
\end{equation}
which concludes the proof of the lemma.

\section{Proof of Lemma 3}
\renewcommand{\theequation}{\thesection.\arabic{equation}}
The average probability of error at high SNR is given by
\begin{equation}
\label{Appp3_3}
P_{e}(\lambda) \asymp P(\mathcal{H}_{o}) \frac{\Gamma(M, \frac{\lambda}{2})}{\Gamma(M)} + P(\mathcal{H}_{1}) \frac{\lambda}{2\overline{\gamma}(M-1)}.
\end{equation}
Through the second derivative test, it can be easily shown that $P_{e}(\lambda)$ is concave for $\lambda < 2M$ and convex elsewhere. Thus, $P_{e}(\lambda)$ has one maximum at $\lambda_{max}$ and one minimum at $\lambda_{min}$. The optimum threshold is $\lambda_{min}$ and is greater than $\lambda_{max}$. The maximum and minimum of $P_{e}$ are obtained by equating $\frac{d P_{e}}{d \lambda}$ to zero
\begin{equation}
\label{Appp5_3}
P(\mathcal{H}_{o}) \frac{-e^{\frac{-\lambda}{2}} \lambda^{M-1}}{2^{M-1} \Gamma(M)} + P(\mathcal{H}_{1}) \frac{1}{2\overline{\gamma}(M-1)} = 0.
\end{equation}
The solutions of the transcendental Eq. in (\ref{Appp5_3}) are given by the principal and lower branches of the Lambert W function as \cite{32}
\begin{align}
\lambda_1 &=\mu^{\frac{1}{M-1}}\exp\left(-\mathcal{W}_{-1}\left(\frac{-\mu^{\frac{1}{M-1}}}{2(M-1)}\right)\right), \nonumber \\
\lambda_2 &= \mu^{\frac{1}{M-1}}\exp\left(-\mathcal{W}_{o}\left(\frac{-\mu^{\frac{1}{M-1}}}{2(M-1)}\right)\right), \nonumber
\end{align}
where $\mu = \frac{P(\mathcal{H}_{1})}{P(\mathcal{H}_{o})} \frac{2^{M-2} \Gamma(M-1)}{\overline{\gamma}}$. Given that $-\mathcal{W}_{-1}(x)$ is always greater than $-\mathcal{W}_{o}(x)$ for $x$ $<$ 0, the optimal threshold is simply $\lambda_{opt} = \lambda_1$, which concludes the proof.

\section{Proof of Lemma 4}
\renewcommand{\theequation}{\thesection.\arabic{equation}}
The series expansion of the Lambert W function is given by \cite{32}
 \[\mathcal{W}_{-1} (x) = L_1-L_2+\sum_{\ell=0}^{\infty}\sum_{m=1}^{\infty}\frac{(-1)^{\ell}\left [\begin{matrix} \ell+m \\ \ell + 1 \end{matrix}\right ]}{m!} L_1^{-\ell-m} L_2^{m},\]
where $L_{1} = \log(-x)$ and $L_{2} = \log(-\log(-x))$. As $x \to 0^{-}$, the first two terms dominate and $\mathcal{W}_{-1} (x) \approx \log(-x)-\log(-\log(-x))$. Thus, from Lemma 3, the optimal threshold can be written as
\[\lambda_{opt} = \mu^{\frac{1}{M-1}} \exp(-L_{1}+L_{2}),\]
which can be expanded as
\begin{align}
\lambda_{opt} &\approx
 \mu^{\frac{1}{M-1}}\exp\Bigg(-\log\left(\frac{\mu^{\frac{1}{M-1}}}{2(M-1)}\right) + \nonumber \\ &\log\left(-\log\left(\frac{\mu^{\frac{1}{M-1}}}{2(M-1)}\right) \right)\Bigg) \nonumber \\
 &=  2(M-1) \log\left(\frac{2(M-1)}{\mu^{\frac{1}{M-1}}}\right).
\end{align}
Thus, as $\overline{\gamma} \to \infty$, and assuming that $P(\mathcal{H}_{o}) = P(\mathcal{H}_{1})$, the optimal threshold can be approximated as
\[\lambda_{opt} \approx 2(M-1) \log\bigg(\frac{M-1}{\Gamma(M-1)^{\frac{1}{M-1}}} \overline{\gamma}^{\frac{1}{M-1}}\bigg).\]
The false alarm probability in (2) can be expressed in the series form as $P_{F} = \sum_{i=0}^{M-1} \frac{\lambda^{i}}{2^{i}\Gamma(i+1)}e^{\frac{-\lambda}{2}}$ \cite{6}. At high SNR, the last term in the series representation dominates and $P_{F} \approx \frac{\lambda^{M-1}}{2^{M-1}\Gamma(M)}e^{\frac{-\lambda}{2}}$. By setting $\lambda = \theta\lambda_{opt}$, the asymptotic false alarm probability is given by
\begin{align}
P_{F} &\asymp \frac{1}{\Gamma(M)}\left(\theta(M-1) \log\left(\frac{M-1}{\Gamma(M-1)^{\frac{1}{M-1}}}\overline{\gamma}^{\frac{1}{M-1}}\right)\right)^{M-1} \times \nonumber \\ &\left(\frac{\Gamma(M-1)^{\frac{1}{M-1}}}{(M-1) \overline{\gamma}^{\frac{1}{M-1}}}\right)^{\theta(M-1)} \nonumber
\end{align}
and
\begin{align}
P_{md} \asymp \frac{\theta}{\overline{\gamma}}\log\left(\frac{M-1}{\Gamma(M-1)^{\frac{1}{M-1}}} \overline{\gamma}^{\frac{1}{M-1}}\right).
\end{align}
Recalling the definitions in Section II, it is straightforward to see that $d_{F} = \theta$ and $d_{md} = 1$. Thus, the achieved diversity order is given by
\[d_{e} = \min\{\theta, 1\}.\]

\section{Proof of Lemma 5}
\renewcommand{\theequation}{\thesection.\arabic{equation}}
The likelihood function is given by
\[\Lambda(r_{1},r_{2},\ldots,r_{M}) = \frac{f(r_{1},r_{2},\ldots,r_{M} | \mathcal{H}_{1})}{f(r_{1},r_{2},\ldots,r_{M} | \mathcal{H}_{o})}.\]
Based on the signal model presented in Section II, the joint pdf of the sensed samples under hypotheses $\mathcal{H}_{1}$ and  $\mathcal{H}_{o}$ are
\[f(r_{1},r_{2},\ldots,r_{M} | \mathcal{H}_{1}) = \prod_{i=1}^{M} f(r_{i} | \mathcal{H}_{1}) \]
\begin{equation}
\label{App1_1}
= \prod_{i=1}^{M} \frac{1}{\sqrt{2 \pi (1+\gamma_{i,j})}} e^{-\frac{r_{i}^{2}}{2(1+\gamma_{i,j})}},
\end{equation}
and
\begin{equation}
\label{App1_2}
f(r_{1},r_{2},\ldots,r_{M} | \mathcal{H}_{o}) = \prod_{i=1}^{M} \frac{1}{\sqrt{2 \pi}} e^{-\frac{r_{i}^{2}}{2}}.
\end{equation}
By combining (\ref{App1_1}) and (\ref{App1_2}), the Log Likelihood Ratio (LLR) test reduces to
\begin{equation}
\label{App1_3}
\sum_{i=1}^{M} \frac{\gamma_{i,j}}{1+\gamma_{i,j}} |r_{i}|^{2} \mathop{\gtreqless}_{\mathcal{H}_{0}}^{\mathcal{H}_{1}} \eta.
\end{equation}
Because the factor $\frac{\gamma_{i,j}}{1+\gamma_{i,j}}$ is constant over every $l_{j}$ samples and $j$ varies from 1 to $Q$, we can rewrite the LLR test as
\begin{equation}
\label{App1_5}
\sum_{j=1}^{Q} \frac{\gamma_{j}}{1+\gamma_{j}} Z_{i} \mathop{\gtreqless}_{\mathcal{H}_{0}}^{\mathcal{H}_{1}} \eta,
\end{equation}
where $Z_{j} = \sum_{i=l_{j-1}+1}^{l_{j-1}+l_{j}} |r_{i}|^{2}$ and $l_{o} = 0$. This concludes the proof of the lemma.

\section{Proof of Lemma 6}
\renewcommand{\theequation}{\thesection.\arabic{equation}}
As stated in Proposition 1, the optimum threshold can be obtained by solving the equation $d_{F}(\lambda)$ = $d_{md}(\lambda)$ for $\lambda$. Unlike the NP test, we do not know how $\lambda_{opt}$ affects the diversity order as the functional form of $\lambda_{opt}$ in terms of $\overline{\gamma}$ is unknown. Thus, applying the definition of diversity order in Section II to Eq. (\ref{mixedeq}), we have $d_{md} = \frac{-M \log(\lambda)}{\log(\overline{\gamma})} + \min \{ Q, M \}$, where the factor $\min\{Q, M\}$ results from the fact that if $Q > M$, we can switch the antenna modes $M$ times only. The diversity order at large SNR is given by $\frac{-\log(P_{F})}{\log(\overline{\gamma})}$. Hence, the error probability diversity order is
\begin{equation}
\label{divord}
d_{e} = \min\left\{\frac{-\log(P_{F})}{\log(\overline{\gamma})}, \,\,\, -\frac{M \log(\lambda)}{\log(\overline{\gamma})} + \min\{Q, M\}\right\}.
\end{equation}
From Proposition 1, we need to find $\lambda_{opt}$ that satisfies $d_{md}(\lambda) = d_{F}(\lambda)$, which can be reduced to $\frac{\lambda^{M-1}}{2^{M-1}\Gamma(M)} e^{-\frac{\lambda}{2}} = \lambda^{M} \overline{\gamma}^{\min\{M,Q\}}$. Thus, similar to the solution of the transcendental equation in Appendix D, the optimum threshold is given by the Lambert W function as
\[\lambda_{opt} = 2 \mathcal{W}_{o}\left(\frac{1}{2 \zeta}\right),\]
where $\zeta = \overline{\gamma}^{-\min\{M,Q\}}2^{M-1}\Gamma(M)$. By replacing the Lambert W function with its asymptotic series expansion and considering the dominant terms as shown in Appendix E, the optimum threshold at large SNR is
\begin{equation}
\label{Optth}
\lambda_{opt} \approx 2\log\left(\frac{\overline{\gamma}^{\min\{M,Q\}}}{2^{M} \Gamma(M) \log\left(\frac{\overline{\gamma}^{\min\{M,Q\}}}{2^{M} \Gamma(M)}\right)}\right).
\end{equation}
By substituting $\lambda$ with $\theta \lambda_{opt}$ in the asymptotic expression of $P_{F}$, it is easy to show that $d_{F} = \theta \min\{Q,M\}$. Besides, it is obvious from (\ref{Optth}) that $\lim_{\overline{\gamma} \to \infty} \frac{\log(\lambda_{opt})}{\log(\overline{\gamma})} = 0$. Combining this result with (\ref{divord}), we have $d_{e} = \min\{\theta \min\{M,Q\}, \min\{M,Q\}\}$, which concludes the proof.








\end{document}